\documentclass[fleqn,usenatbib]{mnras} 

\usepackage{newtxtext,newtxmath}

\usepackage[T1]{fontenc}
\usepackage{ae,aecompl}
\usepackage{subcaption}
\usepackage{xcolor}
\usepackage[utf8]{inputenc}

\usepackage{etoolbox}
\makeatletter
\patchcmd\@combinedblfloats{\box\@outputbox}{\unvbox\@outputbox}{}{%
   \errmessage{\noexpand\@combinedblfloats could not be patched}%
}%
 \makeatother

\usepackage{amsmath,amstext}
\usepackage[figure,figure*]{hypcap}
\usepackage{newtxmath} 
\usepackage{comment}

\usepackage{graphicx}	
\usepackage{amsmath}	
\usepackage{amssymb}	

\graphicspath{{./figure/}}

\title[The cavity of CQ Tau]{A dust and gas cavity in the disc around CQ Tau\\ revealed by ALMA}

\author[]{M. Giulia Ubeira Gabellini$^{1,2}$\thanks{E-mail: maria.ubeira@unimi.it},
Anna Miotello$^{2,3}$,
Stefano Facchini$^{2,4}$,
Enrico Ragusa$^{1,13}$,
\newauthor
Giuseppe Lodato$^{1}$,
Leonardo Testi$^{2,5,6}$,
Myriam Benisty$^{7,8}$, 
Simon Bruderer$^{4}$,
\newauthor
Nicol\'as T. Kurtovic$^{8}$,
Sean Andrews$^{9}$,
John Carpenter$^{10}$,
Stuartt A. Corder$^{11,12}$,
\newauthor
Giovanni Dipierro$^{13}$,
Barbara Ercolano$^{6,14}$,
Davide Fedele$^{5}$,
Greta Guidi$^{15}$,
\newauthor
Thomas Henning$^{16}$,
Andrea Isella$^{17}$,
Woojin Kwon$^{18,19}$,
Hendrik Linz$^{16}$,
\newauthor
Melissa McClure$^{20}$,
Laura Perez$^{8}$,
Luca Ricci$^{21}$,
Giovanni Rosotti$^{22}$,
\newauthor
Marco Tazzari$^{22}$,
David Wilner$^{9}$
\\
\\
$^{1}$Dipartimento di Fisica, Universit{\`a} Degli Studi di Milano, Via Celoria, 16, Milano, I-20133, Italy.\\
$^{2}$European Southern Observatory, Karl-Schwarzschild-Str. 2, 85748 Garching, Germany.\\
$^{3}$Leiden Observatory, Leiden University, Niels Bohrweg 2, 2333 CA Leiden, The Netherlands.\\
$^{4}$Max-Planck-Institut f\"ur Extraterrestrische Physik, Giessenbachstra{\ss}e, 85748 Garching, Germany.\\
$^{5}$INAF - Osservatorio Astrofisico di Arcetri, Largo E. Fermi 5, 50125 Firenze, Italy.\\
$^{6}$Excellence Cluster `Universe', Boltzmann-Str. 2, D-85748 Garching, Germany\\
$^{7}$Institut de Planetologie et d'Astrophysique de Grenoble (IPAG), BP 53, 38041 Grenoble Cedex 9, France\\
$^{8}$Departamento de Astronom\'ia, Universidad de Chile, Camino El Observatorio 1515, Las Condes, Santiago, Chile\\
$^{9}$Harvard-Smithsonian Center for Astrophysics, 60 Garden Street, Cambridge, MA 02138, USA\\
$^{10}$Department of Astronomy, California Institute of Technology, MC 249-17, Pasadena, CA 91125, USA\\
$^{11}$Joint ALMA Observatory (JAO), Alonso de Cordova 3107 Vitacura, Santiago de Chile, Chile\\
$^{12}$National Radio Astronomy Observatory, 520 Edgemont Road, Charlottesville, VA 22903-2475, USA\\
$^{13}$Department of Physics and Astronomy, University of Leicester, Leicester LE1 7RH, United Kingdom\\
$^{14}$Universitats-Sternwarte M\"unchen, Scheinerstra{\ss}e 1, D-81679 M\"unchen, Germany\\
$^{15}$ETH Zurich, Institute for Particle Physics and Astrophysics, Wolfgang Pauli Strasse 27, 8093 Zurich, Switzerland\\
$^{16}$Max Planck Institute for Astronomy, K\"onigstuhl 17, 69117 Heidelberg, Germany\\
$^{17}$Department of Physics and Astronomy, Rice University, 6100 Main Street, Houston, TX, 77005, USA\\
$^{18}$Korea Astronomy and Space Science Institute, 776 Daedeokdae-ro, Yuseong-gu, Daejeon 34055, Korea\\
$^{19}$Korea University of Science and Technology, 217 Gajang-ro, Yuseong-gu, Daejeon 34113, Korea\\
$^{20}$University of Amsterdam, Postbus  94249 1090 GE  Amsterdam\\
$^{21}$Department of Physics and Astronomy, California State University, 18111 Nordhoff Street, 91130, Northridge, CA, USA\\
$^{22}$Institute of Astronomy, University of Cambridge, Madingley Road, CB3 0HA Cambridge, United Kingdom}
\date{Accepted 2019 April 16. Received 2019 April 9; in original form 2018 December 5.}
\pubyear{2019}

\begin{document}
\label{firstpage}
\pagerange{\pageref{firstpage}--\pageref{lastpage}}
\maketitle
\begin{abstract}
The combination of high resolution and sensitivity offered by ALMA is revolutionizing our understanding of protoplanetary discs, as their bulk gas and dust distributions can be studied independently. In this paper we present resolved ALMA observations of the continuum emission ($\lambda=1.3$ mm) and CO isotopologues ($^{12}$CO, $^{13}$CO, C$^{18}$O $J=2-1$) integrated intensity from the disc around the nearby ($d = 162$ pc), intermediate mass ($M_{\star}=1.67\,M_{\odot}$) pre-main-sequence star CQ Tau. The data show an inner depression in continuum, and in both $^{13}$CO and C$^{18}$O emission. We employ a thermo-chemical model of the disc reproducing both continuum and gas radial intensity profiles, together with the disc SED. The models show that a gas inner cavity with size between 15 and 25 au is needed to reproduce the data with a density depletion factor between $\sim 10^{-1}$ and $\sim 10^{-3}$. 
The radial profile of the distinct cavity in the dust continuum is described by a Gaussian ring centered at $R_{\rm dust}=53\,$au and with a width of $\sigma=13\,$au. 
Three dimensional gas and dust numerical simulations of a disc with an embedded planet at a separation from the central star of $\sim20\,$au and with a mass of $\sim 6\textrm{--} 9\,M_{\rm Jup}$ reproduce qualitatively the gas and dust profiles of the CQ Tau disc. However, a one planet model appears not to be able to reproduce the dust Gaussian density profile predicted using the thermo-chemical modeling.
\end{abstract}

\begin{keywords}
protoplanetary discs -  astrochemistry - planet-disc interactions - hydrodynamics 
\end{keywords}



\section{Introduction}
Protoplanetary discs are the natural outcome of the star formation process \citep{1987ARA&A..25...23S}. Material infalling from the pre-stellar core is channeled into the central star and, due to angular momentum conservation, eventually forms a disc. Since the earlier stages of this process, planetary systems form from the material present in the circumstellar discs. The structure and evolution of protostellar discs are important keys in understanding the planet formation process. The main scenarios currently considered for planet formation are the core accretion model (e.g. \citealt[][]{Pollack1996}) and disc instability (e.g. \citealt[][]{Boss1997}). 
According to the former scenario, planets form through the sequential aggregation of the solid component present in protoplanetary discs and eventually form planetesimals (e.g. review of \citealt{Testi2014}). 
The embryo accretes through a balance between collision of planetesimals and fragmentation until it has obtained most of its mass \citep{2016SSRv..205...41B}. After this, a rapid gas accretion can occur which leads to the giant planet formation. 
The alternative scenario is related to the development of gravitational instabilities (GI) during the initial stages of the disc evolution. In the outer part of self-gravitating discs, if the disc-star mass ratio is higher than the disc aspect ratio ($M_{\rm disc}/M_{\star} \gtrsim H/R$), the rapid growth of the density perturbations induced by GI may produce bound clumps, although the effectiveness of this model to produce Jupiter mass planets is controversial (e.g. review of \citealt{Kratter-Lodato2016}). Independently of the exact planet formation process, it is clear that grain growth plays an important role in the evolution and dynamics of the disc.

Young massive planets are expected to imprint signatures on the dust and gas structure of their parent protoplanetary disc, such as cavities, gaps or asymmetries, which can be detectable in the scattered light and thermal emission of discs. Thanks to the new facilities, like the Atacama Large Millimeter/submillimeter Array (ALMA), SPHERE on the Very Large Telescope (VLT) and GPI at the Gemini Telescope, such signatures are now observed with relative ease \citep[e.g.][]{Garufi13,Marel13,benisty2015,benisty2017,benisty2018,2015ApJ...808L...3A,2016ApJ...820L..40A,2016Sci...353.1519P,2016PhRvL.117y1101I,Fedele2017,2017A&A...597A..32V,2017arXiv171006485P,Fedele2018,Hendler2018,Dipierro2018,liu18,long18,2018ApJ...866L...6C}. 
Of particular interest are those discs whose emission shows a dip at NIR wavelengths in the Spectral Energy Distribution (SED), indicating a decrease in the NIR grain opacity. Although there is no agreed explanation of what causes the dip in the NIR, possibly this can be related to the presence of a depletion of small dust grain in the inner disc regions.
The flux emission at longer wavelengths resembles the typical emission coming from a dust-rich object in the outer regions \citep[][]{1989AJ.....97.1451S,Calvet2005}. These discs are referred to as transitional discs (TDs) \citep[][]{Espaillat2014}. The term `transitional' indicates that these objects may be in a transition phase from optically thick gas/dust-rich discs extending inward to the stellar surface to objects where the disc has been dispersed. Whether all discs go through such a phase is however still debated. These discs are anyway excellent candidates to test planet formation theories.

Quantifying the amount of dust and gas content inside the cavity of transitional discs is of fundamental importance in order to distinguish between different clearing mechanisms. The main processes invoked so far are:
1) Dispersal by (photoevaporative) winds \citep[e.g.][]{Clarke2001}, 
2) Dynamical clearing by a (sub-)stellar companion \citep[e.g.][]{Papaloizou2007},
 3) MHD winds and dead zones \citep[e.g.][]{2015A&A...574A..68F, Pinilla2016}. 
 Grain growth was considered as another possible mechanism \citep[e.g.][]{Dullemond2005, Ciesla2007, Birnstiel2012a}. However, recently it was found not to alter the gas profile significantly \citep{Bruderer13} and to have difficulties in explaining the large observed cavities at (sub-)mm wavelengths \citep{2014ApJ...780..153B}.
Photoevaporation, dynamical clearing and dead zones are the most accredited scenarios. These processes are not mutually exclusive and can probably operate at the same time \citep[][]{Williams2011, Rosotti2013}. 

One of the main goals of current planet-formation studies is to find planets still embedded in protoplanetary discs, in order to catch the planet formation process as it happens. A potential detection of a planet in discs with cavities may put strong constrains on formation timescales and eventually explain characteristics of observed exoplanets \citep{Simbulan2017}. The statistics of planet candidate detection in cavities of discs is still low, but the capabilities of new instruments are beginning to provide us with some new companion candidates (e.g. \citealt[][]{Quanz2013,2017arXiv171011393R}).  Recently, \citealt[][]{Keppler2018} have detected and confirmed a point source within the gap of the transition disc around PDS 70. 
Considering the difficulties to obtain a secure direct detection, a complementary method consists of inferring the presence of protoplanets by looking at structures in discs, and comparing them with planet-disc hydrodynamical simulations to derive constraints on planetary masses (\citealt[][]{2016ApJ...818...76J, Rosotti2016, 2018ApJ...857...87L, dipierro18}).

Before ALMA, the spatial distribution of gas and dust were considered to be similar, and the mm-continuum data was used to trace the gas content assuming a gas-to-dust ratio $\sim 100$ (similar to the interstellar ratio). 
Recent observations, on the contrary, reveal a discrepancy between the gas and dust disc sizes. The disc emission observed at short wavelengths (or in CO emission lines) is more extended than the disc emission observed at long wavelengths (\citealt[e.g.][]{Tazzari2016}) 
and there is now evidence for the presence of gas inside the dust cavity (\citealt[][]{Andrews2012,Marel13,Bruderer14}).
Moreover, dust grains have probably grown into larger aggregates and their spectral index may even vary across the disc (\citealt[][]{Rodmann2006,Guilloteau2011,Perez12,Birnstiel2012b,Menu2014,Perez2015,Tazzari2016,Liu2017}). This leads to different distributions of micron- and millimeter-sized grains (\citealt[][]{Garufi13,Marel13,Pohl2017,Feldt2017}). 
In order to overcome the uncertainties in the estimate such profiles, a direct measurement of molecular gas is necessary to determine the spatial distribution of gas and how the gas-to-dust ratio varies in protoplanetary discs (\citealt[][]{Bergin2013, Miotello14, Williams2014, Miotello2017, Zhang2017}).

One of the methods currently used to probe the spatial distribution of gas is to detect the line emission of the most abundant molecular species, i.e. CO and its less abundant isotopologues. The dominant constituent of the gaseous disc, H$_{2}$, lacking a permanent electric dipole moment, is difficult to detect; whereas the second most abundant molecule, CO, can be a possible alternative to probe the gas content. $^{12}$CO is the most abundant isotopologue, but its low-$J$ rotational transitions becomes optically thick at low column density. For this reason it is a poor tracer of the gas content, however, its emission lines are a good probe to constrain the temperature at the disc surface. Less abundant isotopologues, such as $^{13}$CO and, in particular, C$^{18}$O have less optically thick lines, which can be used to trace the gas down to the midplane (\citealt[][]{vanZadelhoff2001, Dartois2003}). 
Importantly, optically thin tracers can give a more accurate measure of the gas column density.

The main processes regulating CO abundances, freeze-out and isotope-selective photodissociation, need to be taken into account when interpreting CO isotopologues lines (\citealt[][]{Miotello14, Miotello2016}). Recent ALMA discs surveys \citep[][]{Ansdell2016,Pascucci2016,Miotello2017,Long2017} found very low CO-based gas masses, confirming the discrepancy between CO and HD mass estimates in brighter discs (e.g. \citealt[][]{Bergin2013, McClure2016}). Different processes sequestering elemental carbon may be at play in protoplanetary discs, affecting the CO emission significantly (\citealt[][]{Bruderer12, Bergin2013, Favre13, Miotello2017,2017ApJ...841...39Y,2017ApJ...849..130M}, see Section \ref{dgratio}).\\

CQ Tau is an ideal candidate to perform a comparative analysis between observations and simulations. While it is not a transitional disc by definition (since it has NIR excess), CQ Tau is known to have a cavity at mm wavelengths in the dust continuum \citep{Tripathi2017,2018arXiv180407301P}. In this paper, we report ALMA observations of the system in both continuum and CO integrated intensity at unprecedented angular resolution and model the gas and dust emission to determine the disc density structure. This structure is then compared to global three dimensional hydrodynamical simulations of a disc hosting a planet. The paper is organized as follows: in Section \ref{2} we describe the properties of the target and the ALMA observations. In Section \ref{3} we present the method used for the modeling. In Section \ref{4} we described the modeling results from the physical-chemical code DALI \citep[Dust And LInes,][]{Bruderer12}. { In Section \ref{5} we studied the possibility of the cavity to be cleared by an embedded planet through a set of hydrodynamical simulations which allowed us to derive the planet mass and location. Moreover, we discuss the results from the hydro simulations with the physical-chemical code DALI and compare them with other possible mechanism to open a cavity (Section \ref{6}). Finally we conclude our work with Section \ref{8}.}

\section{Observations}\label{2}
\begin{figure*}[]
\includegraphics[width=0.49\textwidth]{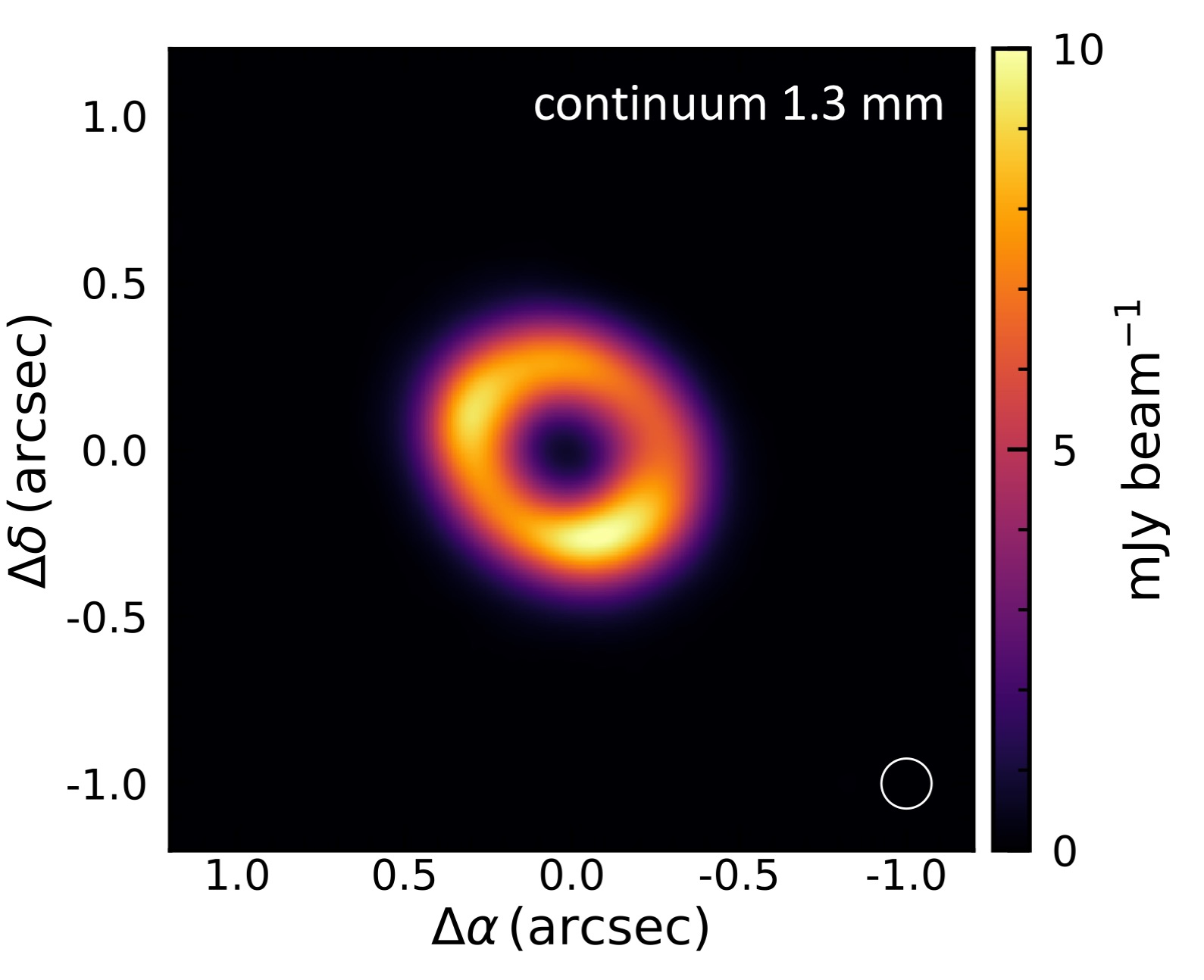}
\includegraphics[width=0.49\textwidth]{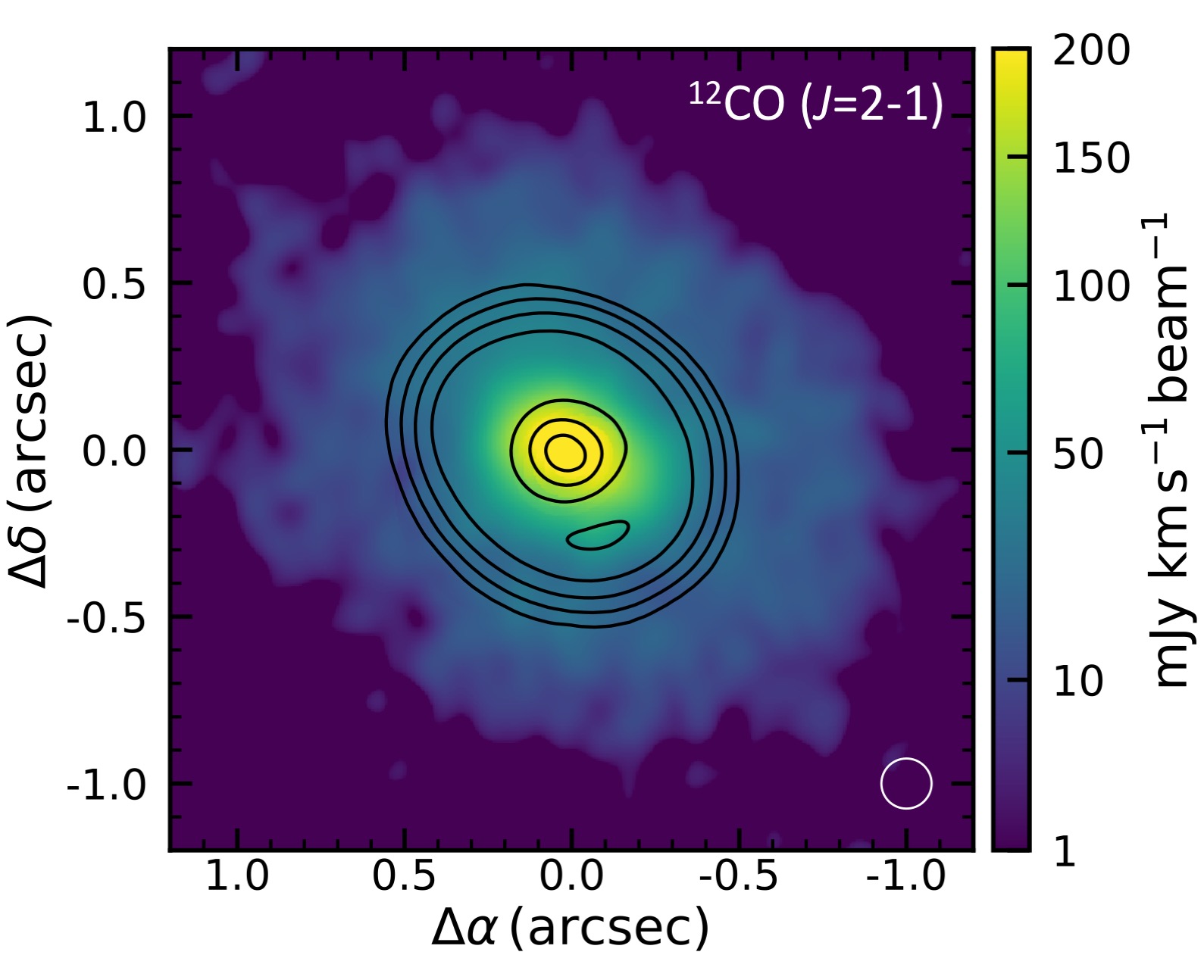}\\
\includegraphics[width=0.49\textwidth]{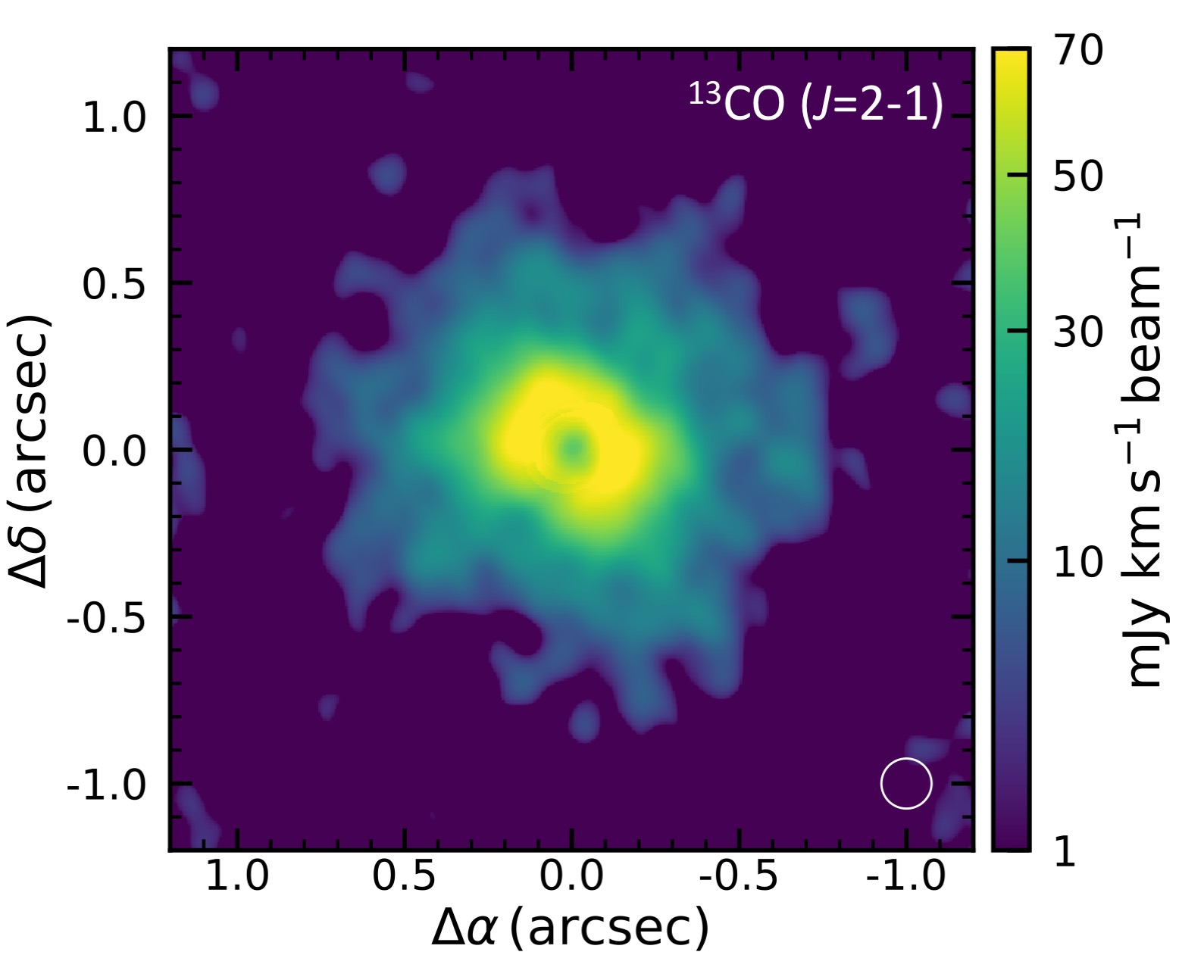}
\includegraphics[width=0.49\textwidth]{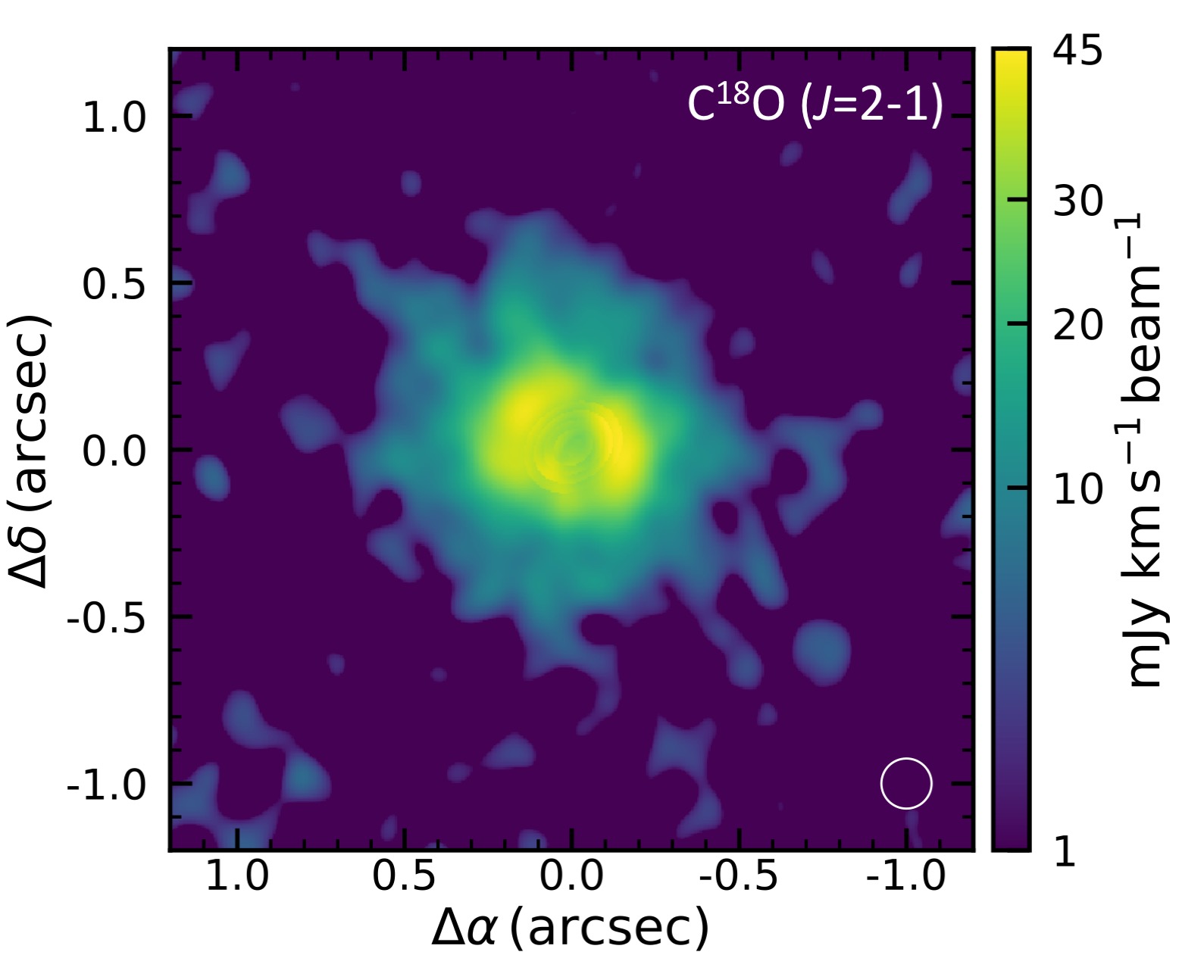}
\caption{ALMA observation of the continuum at 1.3 mm on the upper left panel and zero-moment maps of $^{12}$CO (upper right), $^{13}$CO (lower left) and C$^{18}$O (lower right) $J$=2-1. On top on the $^{12}$CO image we overplot the contours of the continuum respectively at 20, 40, 80, 160, 320 $\sigma$. The white circle in the lower right corner indicates the size of the beam: $0.\!\!^{\prime\prime}15$, corresponding to $\sim 24\,$au.}\label{fig:images_fin}
\end{figure*}

\subsection{Target properties}
CQ Tau is a variable star of the UX Ori class with a spectral type F2 (see Table \ref{tab:stellar_prop}). It is a young ($\sim$10 Myr), nearby (162 pc, \citealt{Gaia2016}), intermediate mass (1.67 $M_{\odot}$) star, surrounded by a massive circumstellar disc (\citealt[][]{Natta2000}). 
The continuum emission at mm-wavelength was observed with different instruments: OVRO interferometer (\citealt{Mannings1997}); the Plateau de Bure interferometer at 2.9 and 1.2\,mm (\citealt{Natta2000}); VLA at 7\,mm and 3.6\,cm (\citealt{Testi2001}). The analysis of the SED at mm-wavelengths allows us to infer that dust has grown up to grains larger than the typical ISM size (\citealt{Testi2001, Testi2003, Chapillon2008}). 
 The disc was already observed with the the IRAM array by \citet{Chapillon2008} who detect a weak disc emission in $^{12}$CO. \citet{Mendigutia2012} gave an upper limit for the accretion rate of CQ Tau equal to $\log(\dot{M}_{\rm acc})<-8.3~M_{\odot}$ yr$^{-1}$, whereas \citet{Donehew2011} measured $\log(\dot{M}_{\rm acc})\sim -7~M_{\odot}$ yr$^{-1}$. This suggests that the CQ Tau disc has an high accretion rate.

\citet{Banzatti2011} presented intermediate resolution observations taken with VLA (1.3 - 3.6 cm), PdBI (2.7 - 1.3 mm) and SMA (0.87 mm). They estimate the dust opacity index to be $\beta \sim 0.6 \pm 0.1$, assuming that the opacity follows a power-law function of the wavelength $\lambda$: $\kappa_{\lambda}\varpropto \lambda^{-\beta}$. Combining these new data with previous measurements, the authors were able to probe the significant grain growth occurring in the disc, with the largest grains growing up to $\sim$cm size. \citet{Trotta2013} found also an indication of grain growth variation with radius, where larger grains were found in the inner disc, compared to the outer disc. 
In a recent work by \citet{Tripathi2017}, the authors detected a cavity in the continuum intensity profile of CQ Tau in SMA archival data.
 \citet{2018arXiv180407301P} presented CQ Tau observations with ALMA finding the presence of a dust cavity in the continuum. The disc was also observed in scattered light (PDI) by Benisty et al. (2019, in prep.).

\begin{table}
\centering
\caption{Stellar properties of CQ Tau.}
\label{tab:stellar_prop}
\begin{tabular}{|cccccccc|}
\hline
$^{1}SpT$ & $^{5}L_{*}$ & $^{5}M_{*}$ & $^{5}$age & $^{2}T_{\rm eff}$ & $^{3}d$ & $^{4}\sigma_{d}$ & $^{6}A_{\rm V}$\\
& ($L_{\odot}$) & ($M_{\odot}$) & Myr & (K) & (pc) & (pc) & (mag)\\
\hline
F2 & 10 & 1.67 & 10 & 6900 & 162 & 2 & 1.9\\
\hline
\end{tabular}\\
References: 1. \citealt[][]{Herbig1960,Natta2001} 2. \citealt{Testi2001} 3. \citealt{Gaia2016} 4. \citealt{2018AJ....156...58B} 5. The luminosity L=6.6 L$_{\odot}$ from \citealt{Lopez-2006} (d=130 pc) is re-scaled considering the new distance from GAIA; the value of mass and age considering the new luminosity where derived using the tracks of \citet{Siess2000}. 6. The extinction was measured by fitting the SED shown in Fig. \ref{fig:SED}.
\end{table}

\subsection{Data}
\subsubsection{Observational Strategy and Data Reduction}

We present Band 6 ALMA data of CQ~Tau (RA = 05$^{\rm h}$35$^{\rm m}$58.46712$^{\rm s}$; dec = +24$^{\circ}$44$^{\prime}$54.0864$^{\prime \prime}$) from three separate programs executed during Cycles~2, 4, and 5 (2013.1.00498.S, PI: L.~P\'erez, 2016.A.00026.S, 2017.1.01404.S, PI: L.~Testi). The ALMA correlator was configured to observe simultaneously CO(2--1), $^{13}$CO(2--1), and C$^{18}$O(2--1), as well as the nearby continuum at 1.3 mm. The array configurations used for Cycles~2, 4 and 5 are respectively C34.6, C40.7 and C43.8. 

Data calibration was performed by ALMA following the standard Quality Assessment procedures for each of the three programmes. After downloading the datasets, we reapplied the calibration tables and extracted the CQ~Tau data from each of the four datasets (the Cycle 5 observations included two separate executions). We corrected the data for the known proper motion of the star and performed one combined iteration of phase only self calibration. The combined datasets were then imaged to produce the continuum and integrated line intensity images used in this paper. For the purpose of the analysis presented in this paper, we imaged the data using a Briggs weighting with a robust parameter of +0.5 and a Gaussian restoring beam of $0.\!\!^{\prime\prime}15$, corresponding to a spatial resolution of $\sim$24$\,$au.
This choice was made to maximize the sensitivity to the line emission and for the purpose of deriving average radial intensity profiles of the disc emission. A detailed analysis of the disc at full angular and spectral resolution will be presented in a forthcoming paper. The observation set used in this paper is shown in Figure~\ref{fig:images_fin}, the noise levels achieved are: 30~$\mu$Jy beam$^{-1}$ in the continuum, and 5~mJy beam$^{-1}$~km s$^{-1}$ in the line integrated intensity maps.
We expect the intensity scale of these observations to be accurate within $\sim$~10\%, as described in the ALMA Technical Handbook\footnote{https://almascience.eso.org/documents-and-tools/cycle5/alma-technical-handbook}.

The line emission from the disc shows a clear rotational pattern dominated by the Keplerian motion around the central star. To compute the integrated intensity maps analyzed in this paper, we first compensate for the Keplerian velocity pattern, then we integrated the emission in each spatial pixel using an optimized range of velocities. The procedure we followed is very similar to the one presented in \cite{2018arXiv180406272Y} and~\citet{Ansdell2018}. The parameters used for the Keplerian motion subtraction compensation are: stellar mass M$_\star=1.67$~M$_\odot$, distance from the Sun d=162~pc, position angle of the disc major axis pa=55~deg, and inclination i=35~deg, which optimize the subtraction of the Keplerian pattern for our dataset and are consistent with the values derived by \citet{Chapillon2008} (noting the different definition of the position angle, complementary to ours).

\subsubsection{Observational results} 

\begin{figure}[]
\begin{center}
\includegraphics[width=85 mm,angle=0]{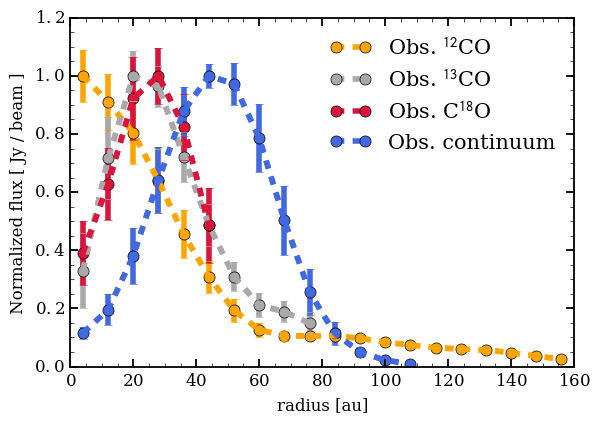}
\caption{Observational profiles done with a radial cut on the major axis of the continuum (blue), $^{12}$CO (orange), $^{13}$CO (gray) and C$^{18}$O (red). Each profile is normalized to the peak value. The lines are derived from the moment zero maps.} \label{fig:PROFILES}
\end{center}
\end{figure}

The map of the continuum emission (upper left of Fig. \ref{fig:images_fin}) presents a compact disc with a clear cavity.
The integrated intensity $^{12}$CO map (upper right of Fig. \ref{fig:images_fin}) reveals the presence of gas inside the cavity seen in the continuum. The emission is centrally concentrated and the majority of the intensity comes from the inner part of the disc. Considering that we expect $^{12}$CO emission to be optically thick, the increase in emission in the inner regions of the cavity is likely due to higher temperatures, and not directly related to the gas surface density (this will be better quantified with the models described in the following section). The $^{13}$CO integrated intensity map (lower left panel of Fig. \ref{fig:images_fin}) shows a decrease of emission in the inner regions of the cavity. The C$^{18}$O integrated intensity map (lower right panel of Fig. \ref{fig:images_fin}) shows a cavity at the same location as in $^{13}$CO.
The continuum and CO isotoplogues radial intensity profiles, shown in Figure \ref{fig:PROFILES}, are obtained with a radial cut on the major axis. The difference between the profiles of the three CO isotopologues can most likely be explained by the different optical depths (e.g. \citealt{Fedele2017}). 
The error bars shown in Figures \ref{fig:PROFILES} are computed as the dispersion of the measured points inside the bin. 

Focusing on the outer disc, the continuum extends radially up to $\sim 0. \, \!\!^{\prime\prime}5$ in radius, while the $^{12}$CO emission extends up to $\sim 1 \, \!\!^{\prime\prime}$. As it was already shown for other discs (\citealt[][]{Panic2009, Andrews2012, Rosenfeld2013, Canovas2016}, \citealt[][]{Ansdell2018}), the gas disc extent on average seems to be a factor of two larger in radius with respect to the mm dust discs.
The emission shows azimuthal asymmetries, both in the continuum and lines. For the purpose of this study, we limit our analysis to the general properties. A follow-up paper will be focused on them in details.

\section{Method}\label{3}
The aim of this work is to obtain a physical-chemical disc model that can simultaneously reproduce the dust and gas emission properties, with a particular focus on the disc cavity. 
In addition to the thermal continuum and CO isotopologues maps previously described, we additionally recover information from the SED, with data taken from previous studies, as described in the Appendix (Table \ref{tab:sed1}). 

\subsection{DALI}\label{dali}
In order to determine the gas and dust temperature and molecular abundances at each position in the disc, we use the physical-chemical code DALI \citep[Dust And LInes,][]{Bruderer12, Bruderer13} that self-consistently computes the physical, thermal and chemical disc structure. 
Given a density structure and a stellar spectrum as inputs, the continuum radiative transfer is solved using a Monte Carlo method to calculate the dust temperature $T_{\rm dust}$ and local continuum radiation field from UV to mm wavelengths. The chemical composition of the gas is then obtained by a chemical network calculation of all the cells. The chemical abundances enter a non-LTE excitation calculation of the main atoms and molecules. The gas temperature $T_{\rm gas}$ is then obtained from a balance between heating and cooling processes. Since both the chemistry and the molecular excitation depend on $T_{\rm gas}$ and vice versa, the problem is solved iteratively. The code is 2D in $R$-$z$. Finally, spectral image cubes are created with a ray tracer fixing the inclination of the disc in the sky at 33$^{\circ}$, assuming perfect azimuthal symmetry. Note that in the DALI modelling the hydrostatic equilibrium is not solved.

\begin{table*}
	\centering
	\caption{\emph{Model parameters}. The fourth column reports the explored range of values for each parameter and the best representative values are presented in boldface.}
	\label{tab:dali_new}
	\begin{tabular}{||llll||} 
    	\hline
        \hline
        & Species & Parameter & Range \\
        \hline
        \emph{Vertical structure}& &$h_{\rm c}$ [rad] & 0.05; 0.07; 0.075; 0.1; \textbf{0.125}; 0.15; 0.2\\
        & & $\psi$  &  0; \textbf{0.05}; 0.1; 0.15; 0.2; 0.25; 0.3\\
        \hline
        \emph{Radial structure}  && $R_{\rm 0}$ [au] & 20; 15; 25; 30; 35; 40; \textbf{56}; 60; 100; 200\\
        &  &$R_{\rm sub}$ [au] & \textbf{0.2}\\
       Self similar density profile (Eq.~\ref{eq:1}) & gas &$\Sigma_{\rm 0, gas}$ [g cm$^{-2}$] & 1; 2; \textbf{2.5}; 5; 7.5; 10; 12.5; 15; 20; 25; 30; 38; 50; 60\\
      Self similar density profile (Eq.~\ref{eq:1}) & small dust & $\Sigma_{\rm 0, small}$ [g cm$^{-2}$] & 0.375; \textbf{0.0375}; 0.00375\\
        & gas,small dust &$R_{\rm cav}$ [au] & 5; 10; 15; \textbf{20}; 25\\
         & gas,small dust &$\gamma_{\rm gas}$ & 0; \textbf{0.3}; 0.5; 0.7; 1.; 1.5\\
        & gas,small dust &$\delta_{\rm gas}$ & 10$^{-4}$; 10$^{-3}$; \textbf{10$^{-2}$}; 5$\cdot 10^{-2}$; 10$^{-1}$; 1; 10$^{1}$\\ 
       Gaussian ring (Eq.~\ref{eq:gaus}) & large dust  & $R_{\rm dust}$ [au] & 52; \textbf{53}; 55; 60\\
       & large dust & $\sigma$ [au] & 10; 11; 12; \textbf{13}; 14; 15; 20; 25\\
       & large dust & $\Sigma_{\rm 0,dust}$ [au] & \textbf{0.6}\\
        \hline
        \emph{Dust properties} &  &Size small grains & \textbf{0.005$\micron$ - 1$\micron$}\\
        & & Size large grains & \textbf{1$ \micron$ - 1cm}\\
        & &  $q_{\rm small}$  & \textbf{3.5}\\
        & & $q_{\rm large}$ & \textbf{3}\\
        & & $\chi$ & \textbf{0.2}\\
        \hline
        \hline
   \end{tabular}
\end{table*}

\subsection{Gas and dust profiles}
We adopt two different surface density radial distributions for the gas and for the mm-sized dust components (see Fig. \ref{fig:sigma}) to reproduce the observations shown in Fig. \ref{fig:PROFILES}. 
For the gas, the density structure follows the simple parametric prescription proposed by \citet[][]{Andrews2011}, which consists in a self-similar solution of viscous accretion disc models \citep[][]{Lynden-Bell-Pringle1974,Hartmann1998}:
\begin{equation}\label{eq:1}
\Sigma(R) = \Sigma_{\rm 0}~\Bigg(\frac{R}{R_{\rm 0}} \Bigg)^{-\gamma} \exp \Bigg[ - \Bigg( \frac{R}{R_{\rm 0}}\Bigg)^{2 - \gamma} \Bigg]
\end{equation}
where $R_{\rm 0}$ is the disc characteristic radius, $\Sigma_{\rm 0} $ the surface density normalization ($\Sigma (R_{\rm 0})=\Sigma_{\rm 0}/e$) and $\gamma$ is the power law index of the surface-density profile.
The millimeter dust emission can in principle be modeled with the same prescription (Eq. \ref{eq:1}, see Section \ref{selfsim_dust} for further explanation), but CQ Tau shows a clear ring-like shaped continuum which can be more naturally described by a Gaussian ring. Furthermore, also \citet{2018arXiv180407301P} have recently modeled CQ Tau lower angular resolution continuum observations with an asymmetric Gaussian profile.
We therefore model the large grains density profile with a Gaussian radial profile:
\begin{equation}\label{eq:gaus}
\Sigma_{\rm dust}(R) = \Sigma_{\rm 0,dust} \exp \Bigg[\frac{-(R-R_{\rm dust})^2}{2\sigma^2}\Bigg]
\end{equation}
where $\Sigma_{\rm 0,dust}$ is the maximum value of the density distribution, $R_{\rm dust}$ is the position of its center and $\sigma$ is its width. 

Two different dust populations are considered into DALI, not only large (1 $\micron$ - 1 cm) but also small (0.005 - 1 $\micron$) grains. While the radial distribution of large grains is described by Eq. \ref{eq:gaus}, small grains follow the gas distribution presented in Eq. \ref{eq:1}, with a different surface density normalization ($\Sigma_{\rm 0}$), taken equal to $\Sigma_{\rm 0, small}$. The size distribution of dust grains is: $n(a) \propto a^{-q}$, where $q$ is the grain size distribution index, taken as $q=3$ for large grains and as $q=3.5$ for small grains, and the maximum size of grains considered is $a_{\rm max} = 1\,$cm. We choose a value of $q=3$ for large grains after performing a test with a value of 3.5. In order to reproduce the SED profile at mm size grains, indeed, we needed to give more weight to the mm size grains (see Fig. \ref{fig:SED}). 

Both continuum and line ALMA data show the presence of a cavity in the inner regions of the disc as shown in Fig. \ref{fig:images_fin}. The size of the cavity and the amount of depletion is not the same for the dust and gas components. We call $ R_{\rm cav}$ the gas cavity radius, while the dust cavity radius is set by the location of the Gaussian ring peak,  $R_{\rm dust}$. Within $R_{\rm cav}$, the gas (and the small dust) surface density is lowered by a factor of $\delta_{\rm gas}$, whereas the large dust depletion is described by the Gaussian profile of Eq. \ref{eq:gaus}, given a width of $\sigma$ (see Fig. \ref{fig:sigma}). 

For the stellar photosphere we considered a blackbody with $L_{*}=10\,L_{\odot}$ and $T_{*}=6900\,$K (Table \ref{tab:stellar_prop}). We do not consider accretion onto the star, since the accretion luminosity contributes to less than $10\%$ to the total luminosity of the star \citep{Meeus2012} and such amount is mainly just important for the inner disc heating.
The inner radius for both gas and dust in our models is set to 0.2 au, following the definition of the dust sublimation radius: $R_{\rm sub} = 0.07 (L_{*}/L_{\odot})^{1/2}$ au (assuming a dust sublimation temperature of ~1500 K; \citealt[][]{Dullemond2005}). 

The vertical density distribution is taken to be a Gaussian with a scale height $H = R h$, with: 
\begin{equation}
h = h_{\rm c} \Bigg(\frac{R}{R_{\rm 0}} \Bigg)^{\psi}
\end{equation}
where $h_{\rm c}$ and $\psi$ are free parameters.
The infrared excess in the SED is directly related to the parameters chosen for the disc scale height. 
The two different dust populations are assumed not to follow the same scale height distribution. The large grains, that account for the bulk of the dust mass in the disc, are settled in the midplane; on the other hand, the smaller grains are well coupled with the gas and follow the same vertical profile. The small population has a scale height of $H$, whereas large grains have a scale height of $\chi H$, reduced by a factor $\chi =0.2$ (see Fig. 1 of \citealt{Trapman2017}). 
Finally, dust opacities considered follow \citet{2001ApJ...548..296W} for a standard ISM dust composition. The optical constants used to compute the dust opacities are as in \citet{2001ApJ...548..296W} and \citet{2003ApJ...598.1017D} respectively for silicates and graphite and the mass extinction coefficients were calculated using Mie theory with the $\rm miex$ code \citep{Wolf2004}. 

\subsection{Determining the best representative model}
We start our investigation by exploring the parameters listed in Table \ref{tab:dali_new}, taken from the values reported in literature (e.g. \citealt{Testi2003, Chapillon2008, Banzatti2011, Trotta2013}). We do not carry out any $\chi^2$ minimization analysis as the complexity of the DALI disc models does not allow us to cover a wide set of parameter values. Nevertheless, we explore the parameter space starting from a sparse grid of models, which is then refined around the values obtained for the best representative models.
The sub-mm continuum emission helps us to fix the parameters related to the radial distribution of large grains in the disc, the size of the dust cavity and the total dust mass. The temperature can vary, changing the vertical structure of the disc and the position or scale height of the cavity wall, which determines the amount of direct irradiation by the star that the disc can receive. Finally, the mid-IR part of the SED and the CO line radial profiles are sensitive to both density structure and temperature of the disc.
In Table \ref{tab:dali_new} the range of values used for each parameter are listed, where the best representative value is shown in bold face. 

\subsubsection{SED and radial profile of the continuum emission}\label{b}
Our first step is to look for a model that reproduces the SED profile between 0.36 $\micron$ and 3.55 cm. The optical part of the SED (from 0.36 to 3.4 $\micron$) is affected by extinction and it is corrected considering the extinction law of \citet{Cardelli1989}, with $R_{\rm v}=3.1$. In order to match the stellar spectrum, an extinction of $A_{\rm v} = 1.9\,$mag is required, consistent with previous works (e.g. \citealt{Lopez-2006}). In our models the vertical structure ($h_{\rm c}$, $\psi$) 
is the main factor which can introduce a shadowing and lower the amount of exposed disc surface accessible to stellar light \citep[][]{Woitke2016}. 
These are key parameters for the final temperature of the disc and to reproduce the observed NIR and FIR excess. 

In order to describe the mm part of the SED, we vary $R_{\rm dust}$, $\sigma$ and $\Sigma_{\rm 0, dust}$, until we match simultaneously  the radial profile of the ALMA continuum observations and the longer wavelengths points in the SED.

\subsubsection{CO emission radial profiles}\label{c}
The following step is to reproduce the gas profiles of the three CO isotopologues. The CO isotopologues radial profiles are more radially extended than the continuum emission (as it is possible to see from the normalized profiles in Fig. \ref{fig:PROFILES}). As many previous studies suggest, a single surface density profile for both gas and dust is not able to reproduce the continuum emission and the CO integrated intensity maps simultaneously, indicating that optical depth effects are not sufficient to explain the different radial extent \citep[e.g.][]{2017A&A...605A..16F} as it is shown in the DALI modelling described in Section \ref{4}.

\begin{figure}[]
\begin{center}
\includegraphics[width=83 mm,angle=0]{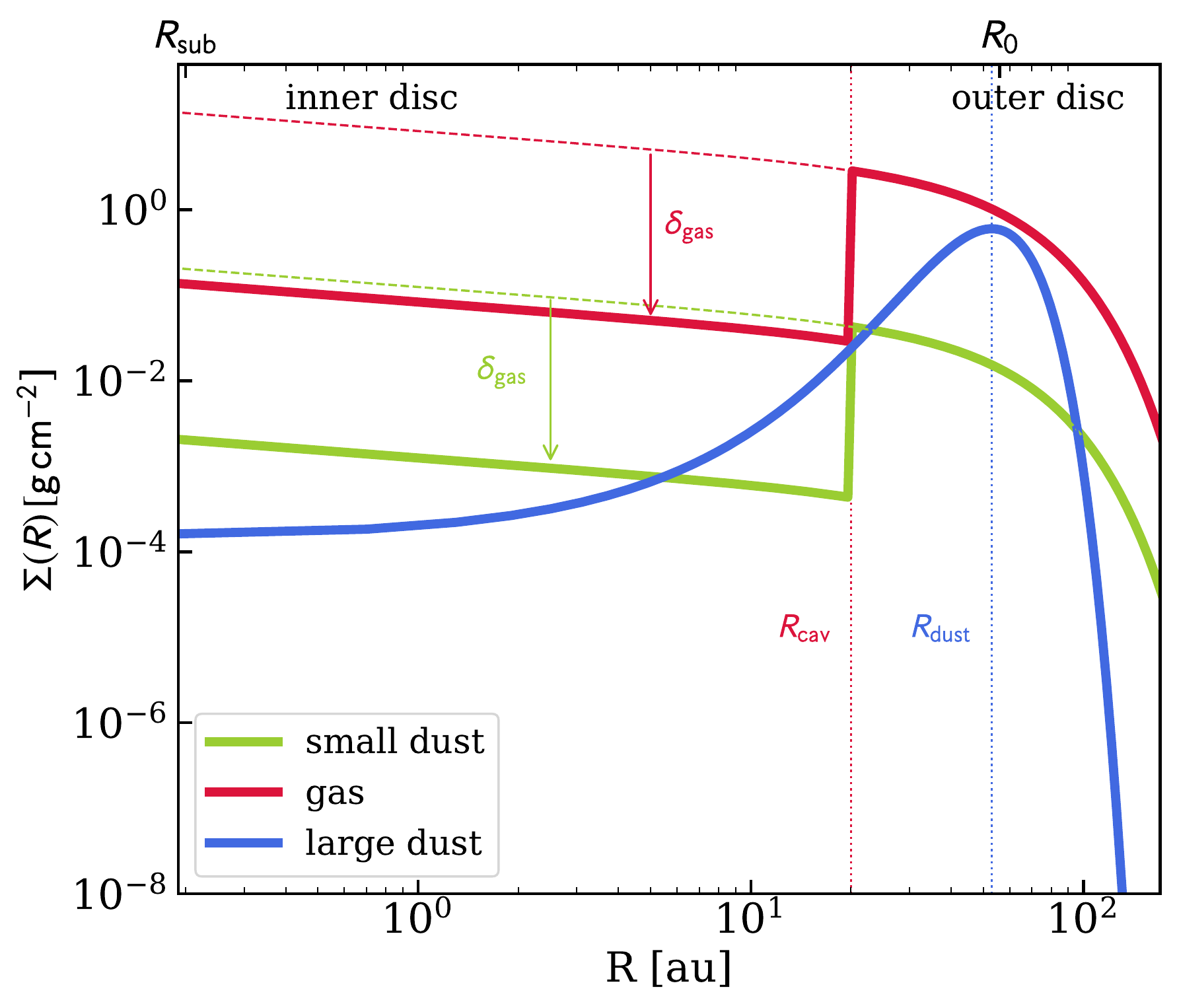}
\caption{Surface density profiles of the best representative model for the three components considered: gas (red line), small grains (green line), and large grains (blue line). The dashed lines show the profiles of the surface density of the disc if the cavity was not present.} \label{fig:sigma}
\end{center}
\end{figure}
\section{Model results}\label{4}
\subsection{SED best representative model}
\begin{figure}[]
\begin{center}
\includegraphics[width=82 mm,angle=0]{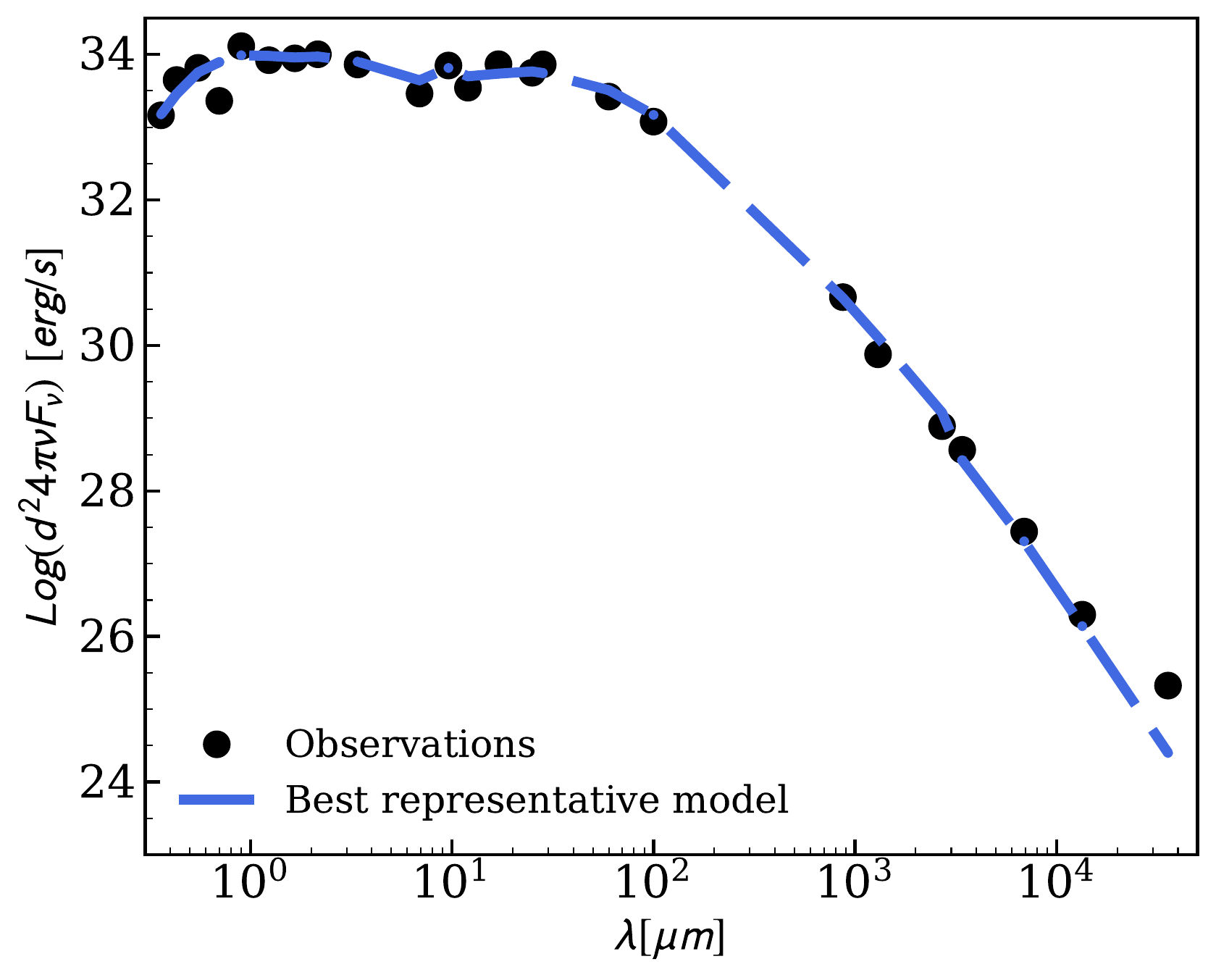}
\caption{Spectral Energy Distribution (SED) profile. The black dots are the observations as described in Table \ref{tab:sed1} in Appendix. The blue dashed line is our best representative model (see Table \ref{tab:dali_new}). 
} \label{fig:SED}
\end{center}
\end{figure}

In order to reproduce the SED profile (Fig. \ref{fig:SED}), we follow the approach described in Section \ref{b}. 
The two main observational constraints to set the best representative model for the SED are the Near-InfraRed (NIR) and Far-InfraRed (FIR) excesses which give information about the vertical structure of the disc, and the total amount of small dust present in the disc. We vary the values of $h_{\rm c}$ and $\psi$, which describe how much the disc is vertically extended and flared. In particular, the parameter which mostly affects the SED is $h_{\rm c}$ \citep{Woitke2016}. Our best representative values for these parameters are $h_{\rm c}=0.125$ and $\psi=0.05$. The range of values under which our models match the observed SED are $0.1<h_{\rm c}<0.15$ and $0.05<\psi<0.1$. We note that a variation in the stellar luminosity value can affect the choice of such parameters.
Longer wavelengths are affected mainly by the total amount of mass present in the disc.
Once the correct amount of total mass is found, we fix this value and vary the other parameters in order to find a good agreement with the ALMA continuum profile. As it is possible to see in Fig. \ref{fig:SED}, the observations are well described by our best representative model. Note that model and data diverge at the wavelength of 3.57 cm. This may be explained by free-free emission, which we do not model, and affects the integrated flux at cm wavelengths. We note also that the SED shows a near-infrared excess together with a resolved cavity in both dust and gas seen in the ALMA data. The small grains are the main responsible for the emission at such wavelengths and a small amount in the cavity is enough to reproduce the NIR emission.

\subsection{Radial profiles}\label{sec:dgradprof}
In order to compare our models with our ALMA observations, we convolve the model results with a Gaussian beam of of $0.\!\!^{\prime\prime}15$, which simulates the resolution of  the observations, both for the continuum and the CO isotopologues. 
We perform a radial cut along the major axis of the continuum and moment zero maps, which has a better resolution compared to an azimuthally averaged profile, in order to compare with our 2D models. 

\subsubsection{A Gaussian ring for large grains}\label{sec:dradprof}
\begin{figure}
\begin{center} 
\includegraphics[width=83 mm,angle=0]{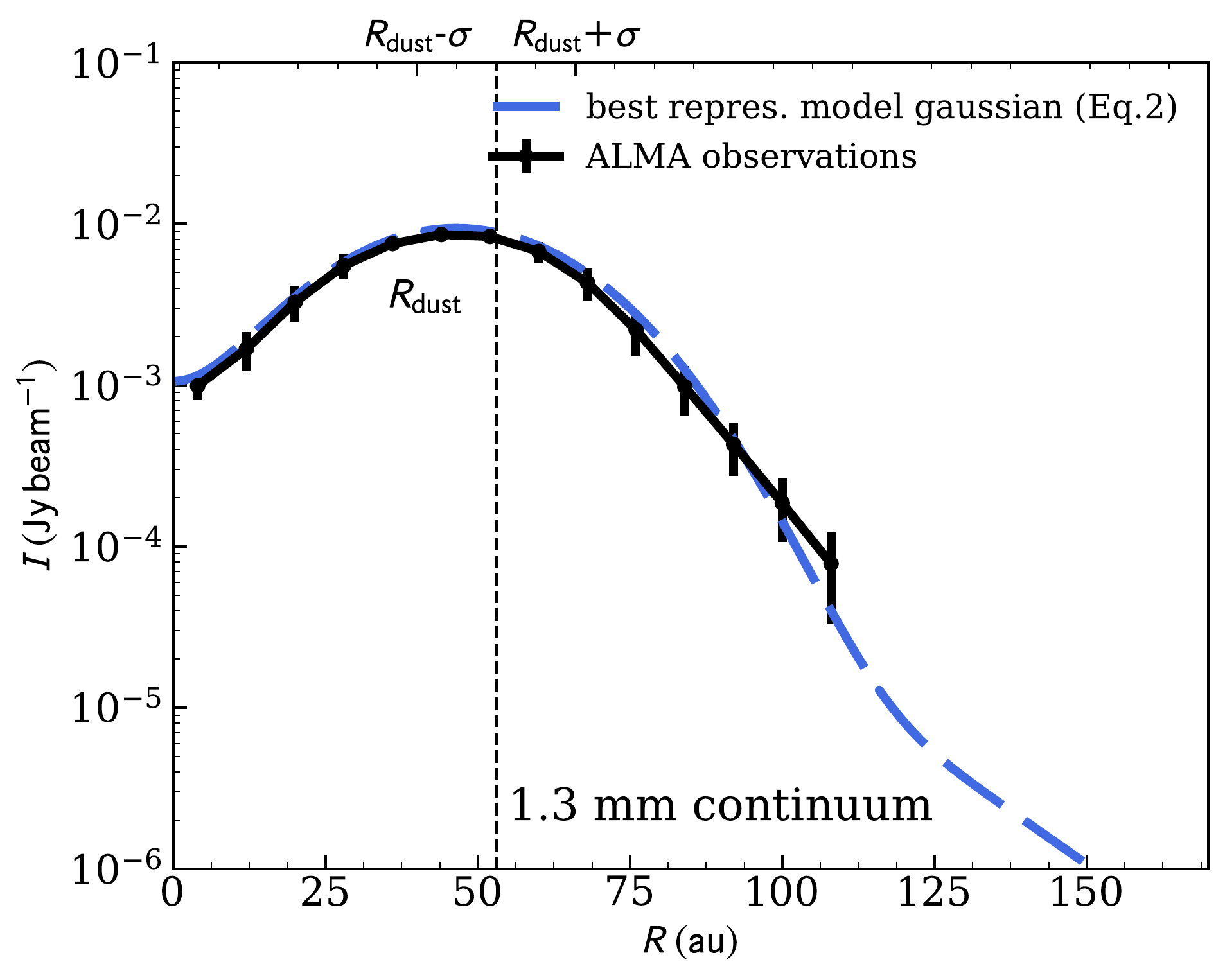}
\caption{Dust continuum taken at 1.3 mm with ALMA. The radial cut on the major axis of observation (black line) is shown along with our best representative model (blue line): Gaussian profile.
}\label{fig:continuumvsR}
\end{center}
\end{figure}

Inspired by the work of \citet{2018arXiv180407301P}, we have employed a Gaussian functional profile, as previously discussed (see Eq. \ref{eq:gaus}). As best representative parameters we find a center peak position of $R_{\rm dust}=53\,$au and a Gaussian width of $\sigma=13\,$au. Our results are consistent with \citet{2018arXiv180407301P} work where the best model had a Gaussian peak radius of $\sim \,$46$\,$au and respectively an inner and outer width of $\sim \,$11$\,$au and $\sim \,$17$\,$au. 
The peak of the Gaussian is $\Sigma_{\rm 0, dust}=0.6\,{\rm g\,cm^{-2}}$. The simulated continuum emission profile is plotted in Fig. \ref{fig:continuumvsR} (blue line) along with the data points (black line). The data are always matched by our model within the error bars, and from now on we will refer to this as the best representative model for the large dust component.
\\
\subsubsection{Gas radial profile}\label{sec:gradprof}
\begin{figure*}[] 
\centering
\begin{subfigure}[t]{0.49\textwidth}
\centering
\includegraphics[width=\textwidth]{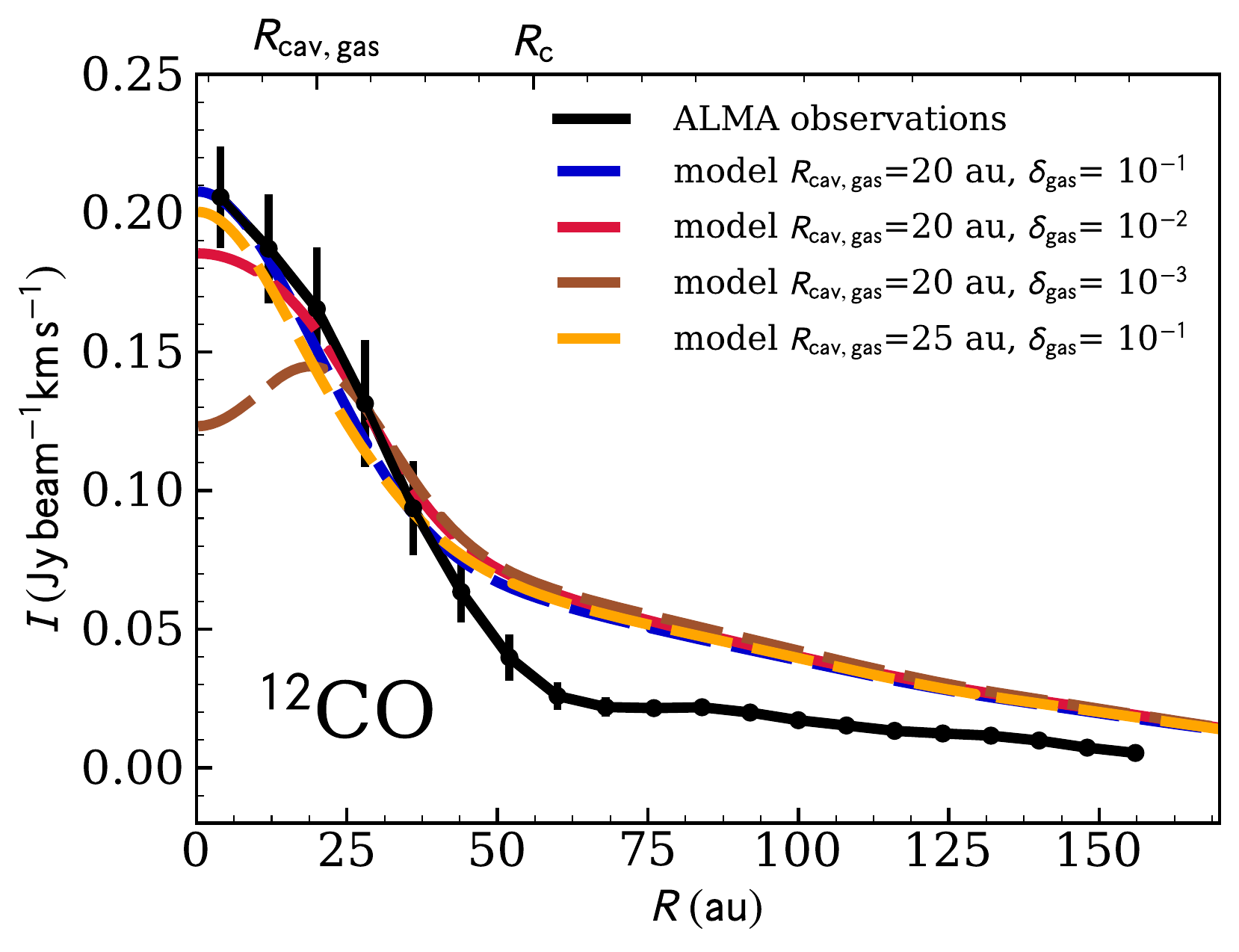}
  \end{subfigure}
  \hspace{0.1 cm}
  \begin{subfigure}[t]{0.49\textwidth}
\centering
\includegraphics[width=\textwidth]{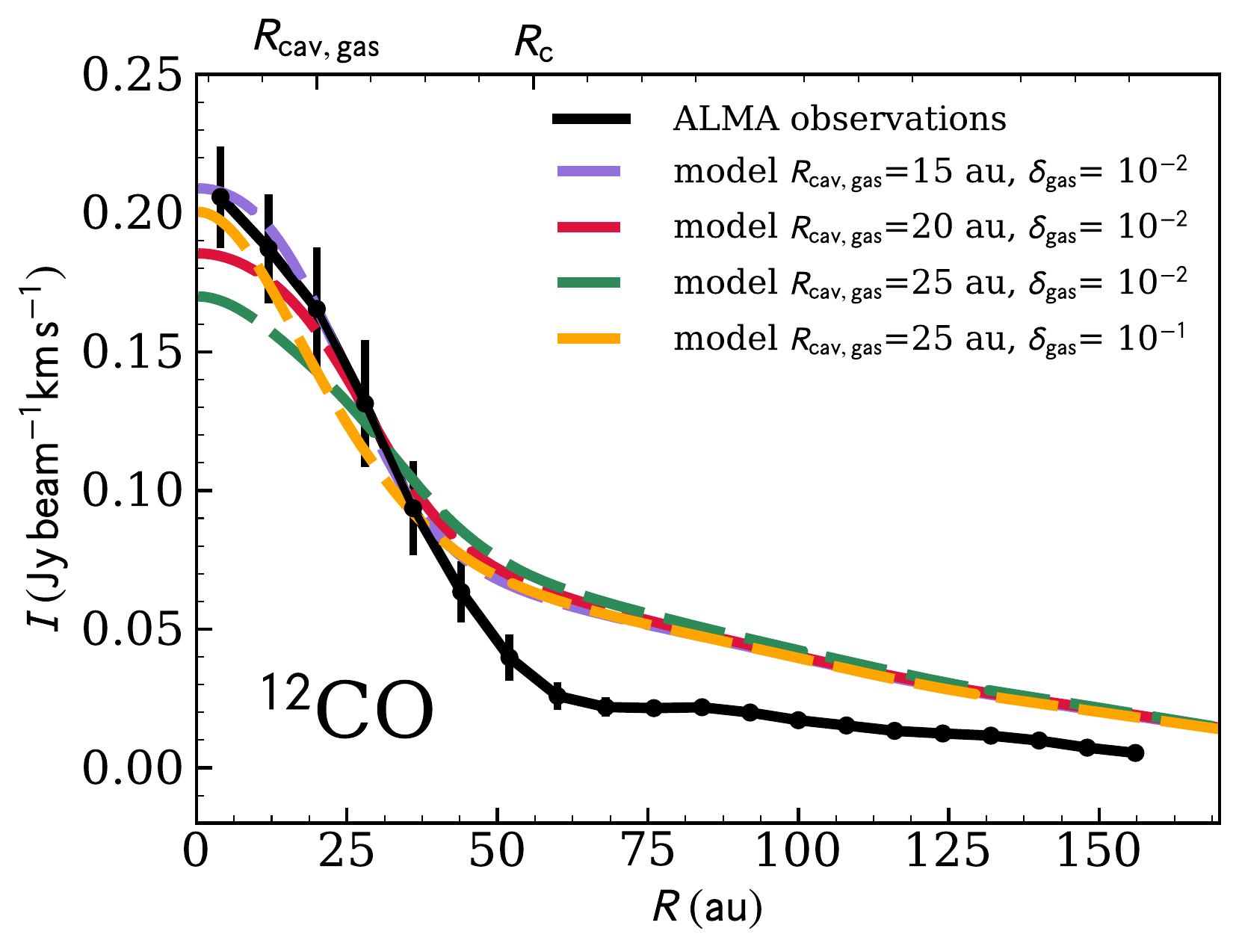}
  \end{subfigure}
   \hspace{0.1 cm}
\begin{subfigure}[t]{0.49\textwidth}
\centering
\includegraphics[width=\textwidth]{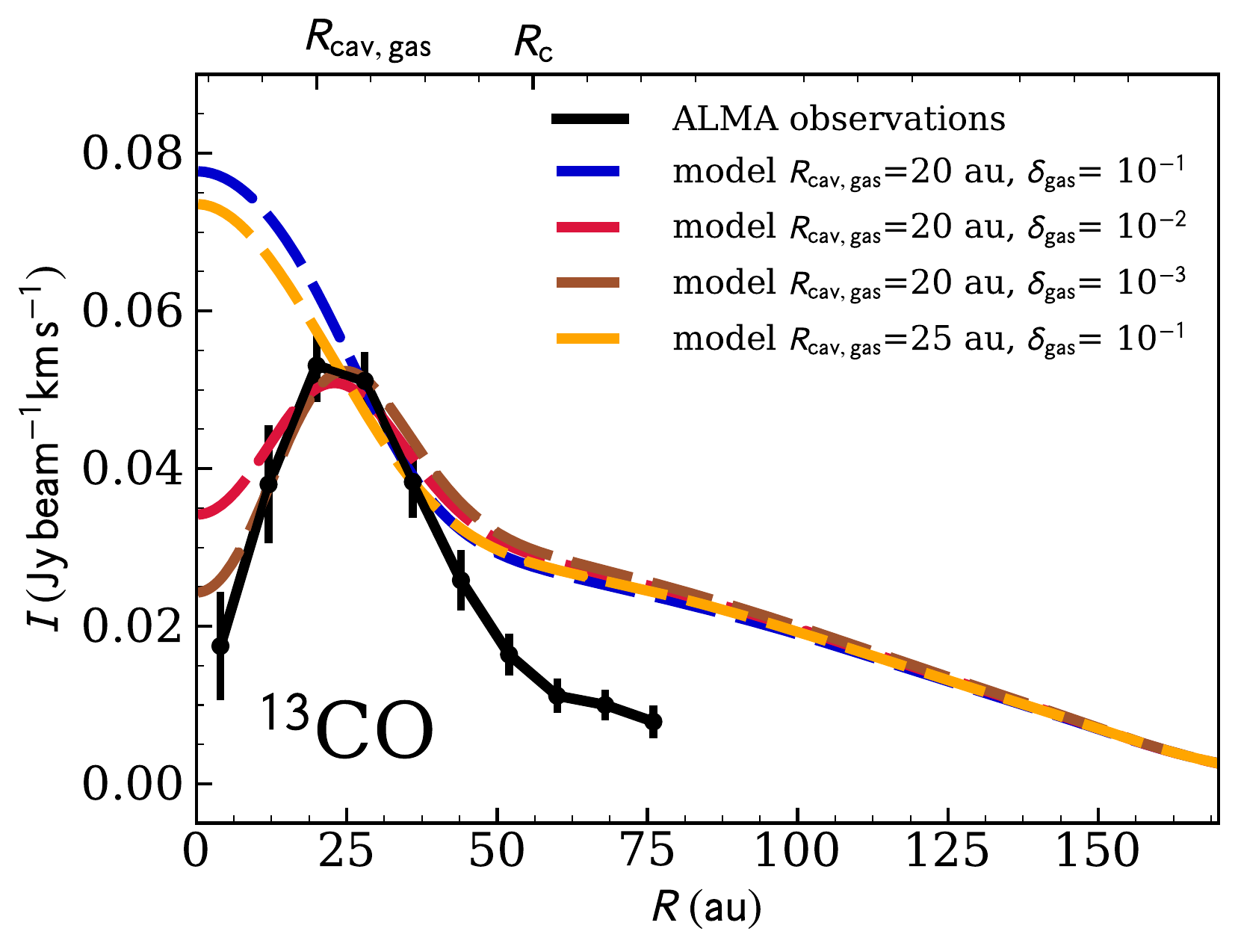}
  \end{subfigure}
  \hspace{0.1 cm}
  \begin{subfigure}[t]{0.49\textwidth}
\centering
\includegraphics[width=\textwidth]{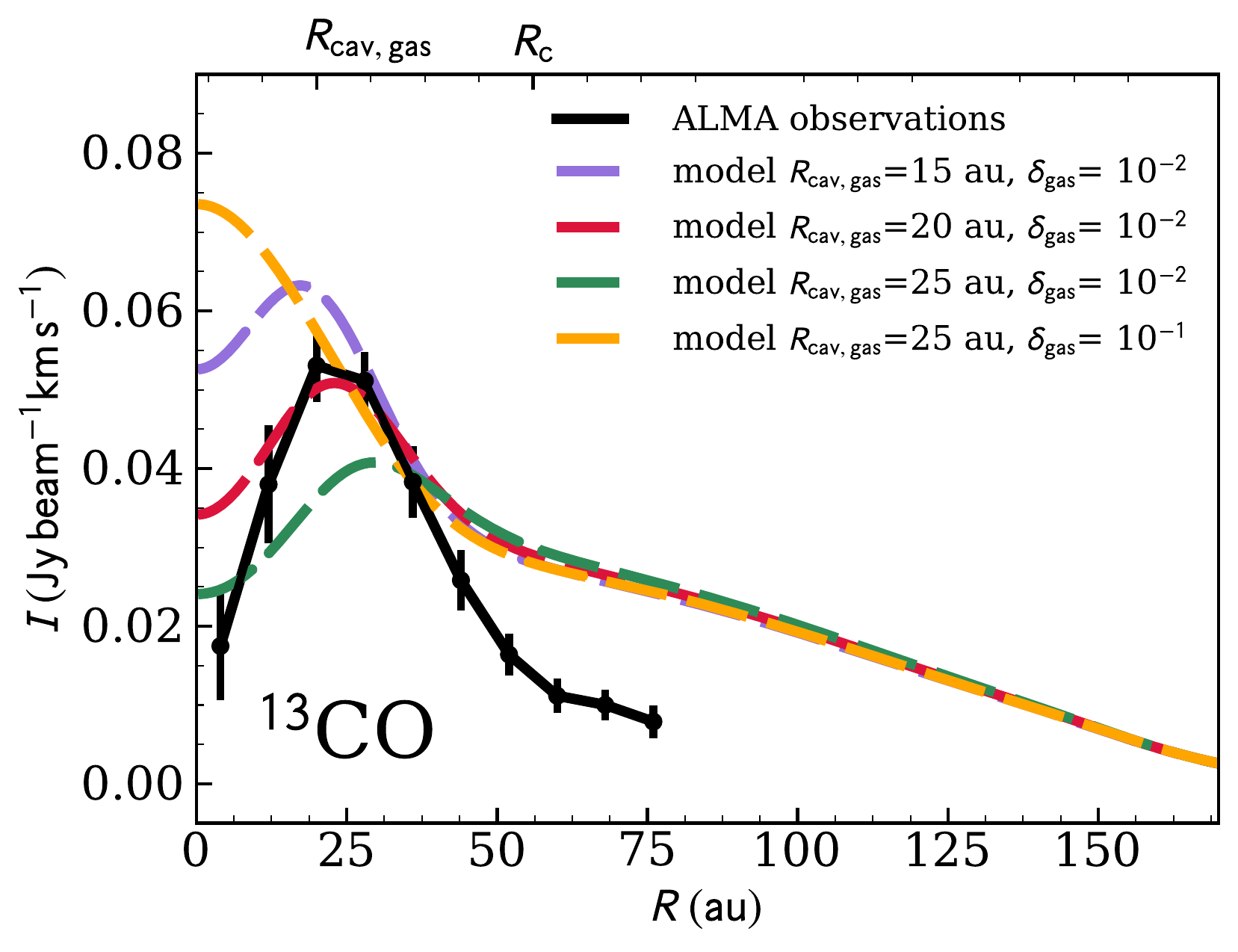}
  \end{subfigure}
    \hspace{0.1 cm}
  \begin{subfigure}[t]{0.49\textwidth}
\centering
\includegraphics[width=\textwidth]{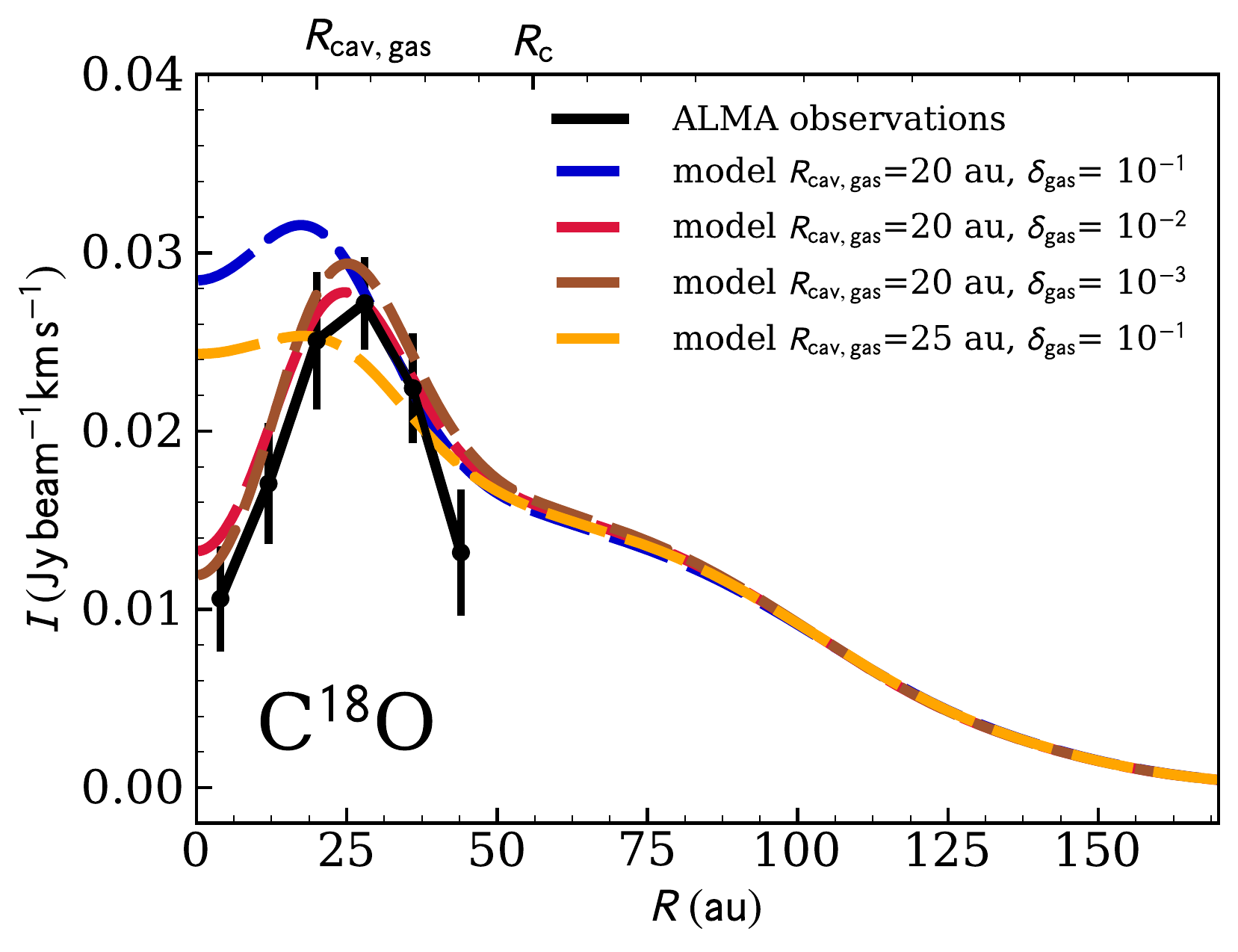}
\end{subfigure}
  \hspace{0.1 cm}
  \begin{subfigure}[t]{0.49\textwidth}
\centering
\includegraphics[width=\textwidth]{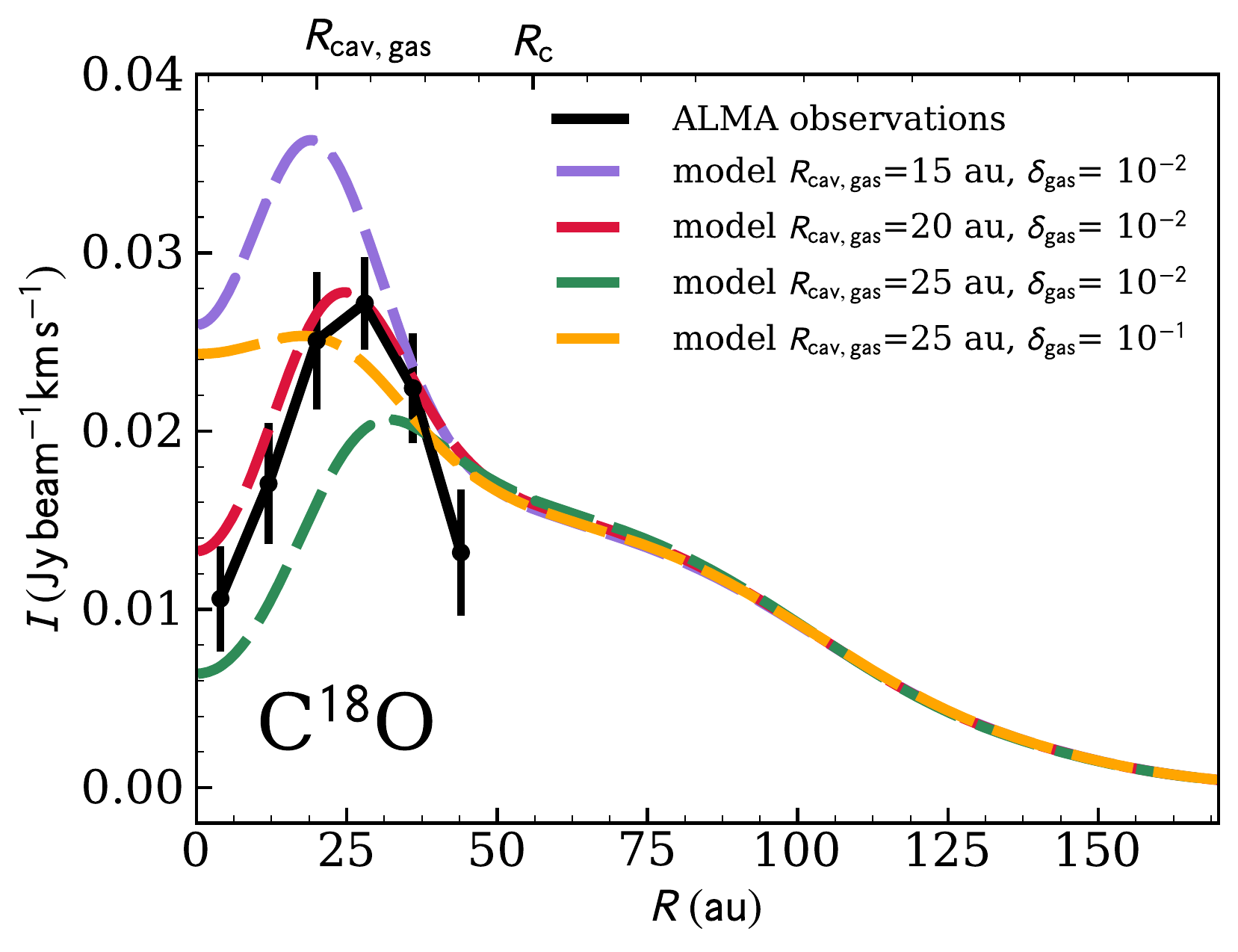}
\label{fig:c18o} 
\end{subfigure}%
\caption{ALMA observations radial cut on the major axis (black line) for gas in the three different isotopologues of CO (top: $^{12}$CO, middle: $^{13}$CO, bottom: C$^{18}$O). Along with our best representative model (red line), on the right we show additional models varying R$_{\rm cav}$, while on the left we vary $\delta_{\rm gas}$. Moreover, we present an additional model (orange), that still describes well the $^{12}$CO profile and it is useful for the model results (see below).
}\label{fig:fig7} 
\end{figure*} 
We use the CO isotopologues observations as a proxy for the gas distribution. In Figs. \ref{fig:fig7} we plot the $^{12}$CO, $^{13}$CO and C$^{18}$O observed radial profiles, along with a selection of our DALI models. In the disc inner regions, where CO lines are more optically thick, the gas temperature plays a big role and strongly depends on the small grains distribution. We implement a relatively high surface density normalization of small grains $\Sigma_{\rm 0, small}=0.0375$ g cm$^{-2}$. 
Setting a higher column density of small grains, while keeping fixed the amount of gas, shifts the dust $\tau$=1 layer upwards and, consequently, a more significant amount of UV photons is absorbed there. The location of the CO isotopologues $\tau$=1 layer does not change significantly but the gas temperature at the emitting layer is reduced as the UV flux is reduced. Moreover, a smaller amount of UV flux can produce a decrease of the dust temperature and consequently, due to thermal balance, can contribute to a lower gas temperature. Accordingly, simulated integrated intensities are lower.
A smaller $\Sigma_{\rm 0, small}$ would lead to higher integrated intensities than what shown in Fig. \ref{fig:fig7}, with a worst match with the observational data. 
All CO isotopologues model results shown in Fig. \ref{fig:fig7} are obtained assuming a gas profile (Eq. \ref{eq:1}) described by the following parameters: $\gamma_{\rm gas}=0.3$, $\Sigma_{\rm 0, gas}=2.5$ g cm$^{-2}$ and a cutoff radius of $R_{\rm 0} = 56$ au. The models always over-predict the emission in the outer disc ($^{13}$CO in particular). Our focus is however on the cavity, as we are interested in constraining which may be the clearing mechanism. Future studies will attempt to give a better description of the CQ Tau outer disc. 

We explored different values for the size and depletion of the cavity (respectively $R_{\rm cav}$ and $\delta_{\rm gas}$) to estimate the best representative model and to gain insights about their degeneracies (Fig. \ref{fig:fig7}).
Observations agreed with a gas depletion factor of 10$^{-3}$ $< \delta_{\rm gas} <$ 10$^{-1}$.
The top left panel of Fig. \ref{fig:fig7} shows that the $^{12}$CO profile cannot be described by a model with a depletion factor of 10$^{-3}$, otherwise the $^{12}$CO emission would also show a depression at small radii. Furthermore, $^{13}$CO and C$^{18}$O profiles suggest that the depletion factor has to be smaller than $\delta_{\rm gas}\lesssim 10^{-1}$ (models shown in blue and yellow do not match the data). Finally, the cavity size is constrained to be larger than 15 au and smaller than 25 au. Our best representative model is shown with a red line in Fig. \ref{fig:fig7} and assumes $R_{\rm cav}=20$ au and $\delta_{\rm gas}=10^{-2}$.  The gas has a smaller cavity radius ($R_{\rm cav}=20\,$au) than the large dust grains ($R_{\rm dust}=53\,$au). The depletion in gas within the cavity is found to be similar to the one of the large grains (see Fig. \ref{fig:sigma}).

\subsection{Dust-to-gas ratio}\label{dgratio}
\begin{figure}
\centering
\includegraphics[width=80 mm,angle=0]{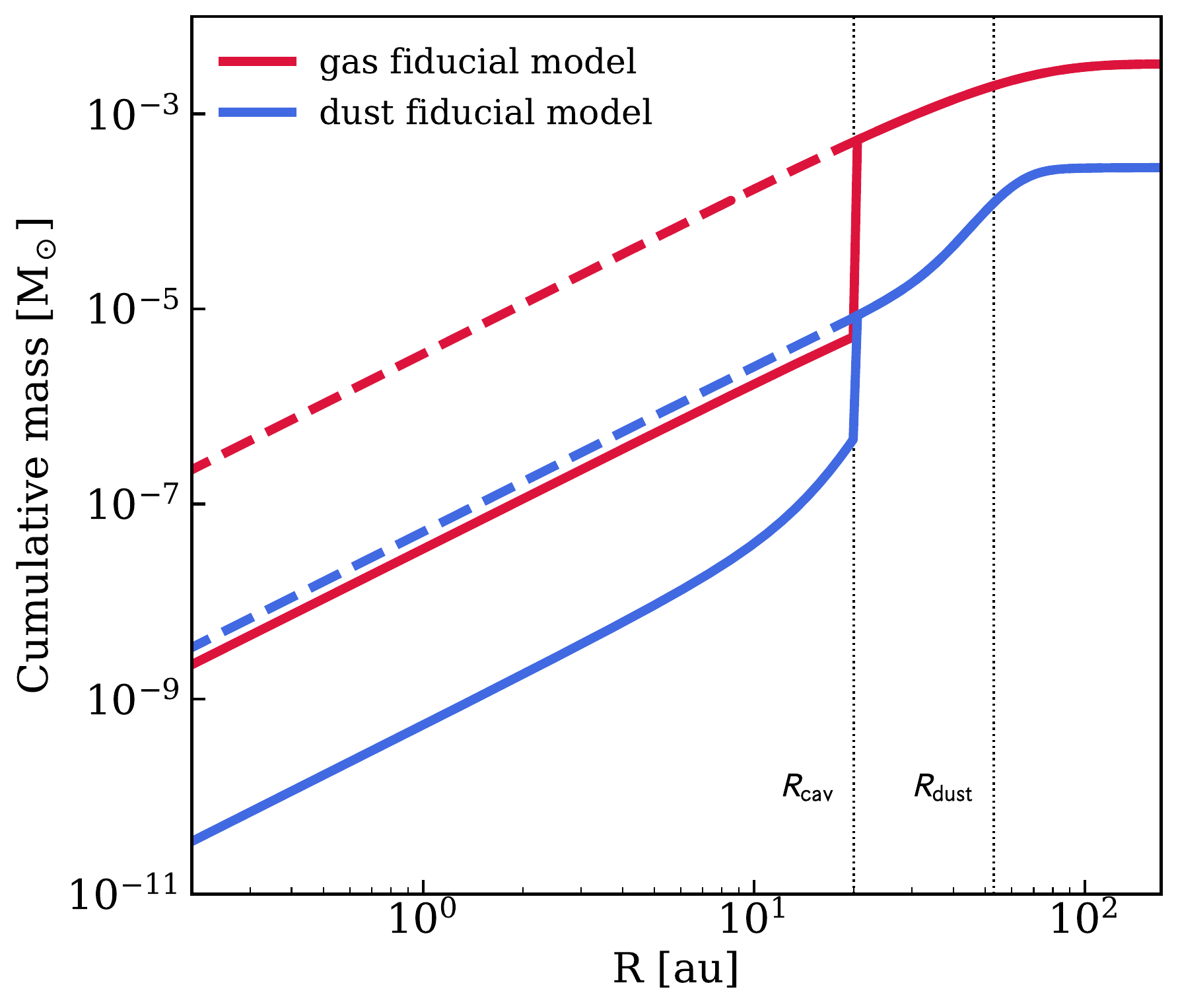}
\caption{Radial profile of the cumulative function of mass of our best representative model (as shown in Fig. \ref{fig:sigma}). The red line shows the gas behavior, while the blue line shows the dust component considering both the small and large grains; the dashed line shows the cumulative mass function without considering the presence of a cavity for both gas (red) and dust (blue). The global dust-to-gas ratio is $\sim 0.09$.
} \label{fig:cumfunct}
\end{figure}
Fig. \ref{fig:cumfunct} shows the cumulative function of mass as a function of the radial distance from the star of both the gas (red) and the dust (blue, small and large grains together) for our representative model. The dust-to-gas ratio is not uniform across the disc, and is close to 0.1.
We compute the total mass for the three different components of our best representative model: $M_{\rm gas} = 3,2 \cdot 10^{-3} M_{\odot}$; $M_{\rm small \, dust} = 4.9 \cdot 10^{-5} M_{\odot}$; $M_{\rm large \, dust} = 2.3 \cdot 10^{-4} M_{\odot}$. The global dust-to-gas ratio for this model of CQ Tau is $\sim 0.09$. This value is higher than the typical ISM values of $\sim 0.01$, but it is consistent with recent findings from ALMA surveys of protoplanetary discs in different star-forming regions (\citealt{Ansdell2016, Ansdell2017, Pascucci2016, Miotello2017, Long2017}). The CO isotopologues emission lines are found to be generally faint compared with the disc continuum emission, resulting in low CO-based gas masses and high CO-based dust-to-gas mass ratios. This may be interpreted as rapid disc physical evolution via loss of gas, or alternatively, as chemical evolution due to carbon depletion processes. The latter may occur if carbon is sequestered from CO and it is either locked into large icy bodies in the midplane, or converted into more complex and less volatile molecules (\citealt{Aikawa1996, Bergin2014, Du2015, Eistrup2016, Kama2016, Yu2016}). 
Current CO and continuum ALMA observations alone cannot distinguish between the two scenarios, but high dust-to-gas mass ratios hint at disc evolution, either physical or chemical. Independent gas mass measurements of discs have been obtained thanks to the detection on hydrogen deuteride (HD) with the PACS instrument on the Herschel Space Telescope in few discs (\citealt[][]{Bergin2013,McClure2016}). When compared with CO isotopologues observations, such data shows that volatile carbon needs to be depleted by different factors depending on the source (\citealt[][]{Favre13,McClure2016}). Unfortunately, no facility is available today to reproduce this kind of observations. Alternatives to constrain the level of volatile carbon and oxygen depletion in discs come from the observations of atomic recombination lines, such as [CI] (\citealt[][]{Kama2016}) or from the detection of hydrocarbon lines (\citealt[][]{Bergin2016,Cleeves2018}).

\subsection{A further test for large dust grains: self similar density profile}\label{selfsim_dust}
In order to model the large dust grains density profile we also attempted to use a self-similar density profile as done for the gas (see Eq. \ref{eq:1}).
\begin{figure}
\begin{center} 
\includegraphics[width=85 mm,angle=0]{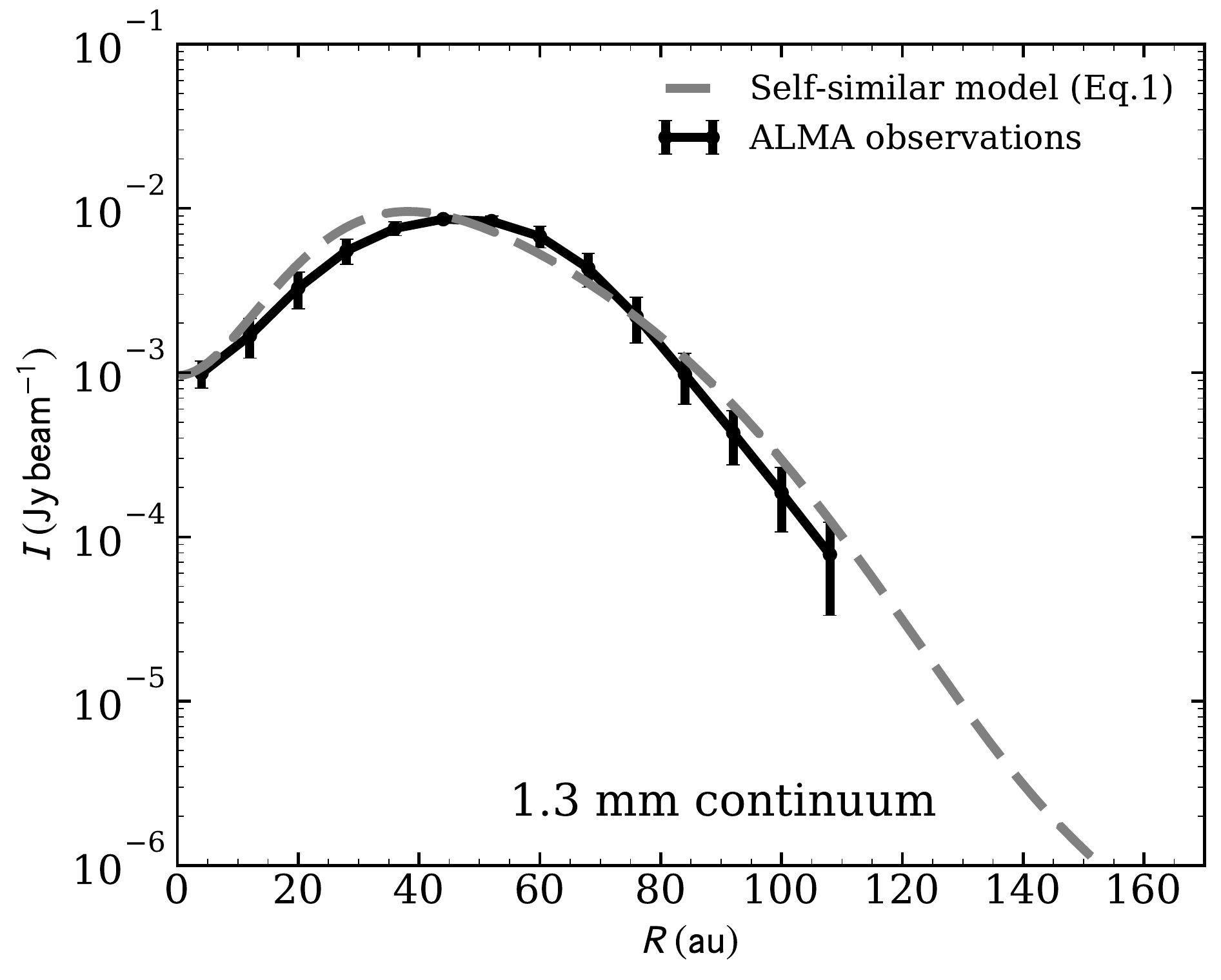}
\caption{Dust continuum taken at 1.3 mm with ALMA. The radial cut on the major axis of observation (black line) is shown along with our best model obtained assuming a self-similar density profile of large dust grains which still can nicely describe our data (see Section \ref{selfsim_dust}). The dust cavity radius is at 28 au and the depletion factor is of $\delta_{\rm dustcav}= 10^{-2}$.}\label{fig:continuumvsR_pl}
\end{center}
\end{figure}
In this case, we used the same cutoff radius as for the gas ($R_{\rm 0}$=56 au), but with a different $\gamma_{\rm dust}$ and surface density normalization of large dust grains $\Sigma_{\rm 0, dust}$, allowed to vary independently. 
The best parameters for large dust grains are found to simultaneously reproduce both the observations of the ALMA continuum radial profile and the SED. The power law index of the surface density distribution is $\gamma_{\rm dust}=-0.7$ and the surface density normalization is $\Sigma_{\rm 0, dust}=0.85$ g cm$^{-2}$. 

To model the inner region of the disc, we consider that the size of the cavity and the amount of depletion can be different between the dust and gas. We choose as dust cavity radius $ R_{\rm cav, dust}$ and as dust cavity depletion $\delta_{\rm dust}$ that are allowed to be different from the gas ones. 
The dust cavity has a radius of 28 au with a depletion of a factor $\delta_{\rm dust}= 10^{-2}$. The proximity of the cavity edge with the cutoff radius implies that the dust is concentrated in a narrow ring.

The self-similar density profile reproduces the observed SED (it overlaps the Gaussian ring SED, blue curve in Fig. \ref{fig:SED}). The resulting continuum emission profile is plotted in gray in Fig. \ref{fig:continuumvsR_pl} along with the data points (black line). From this plot it is possible to see that the self-similar density profile still describes well the large grains behavior. 
\\
\section{Is the cavity induced by an embedded planet?}\label{5}
The cavity visible in the CQ Tau disc can be explained by several mechanisms: dispersal by (photoevaporative) winds, dynamical clearing by a (sub-)stellar companion, MHD winds and dead zones. In this Section, we focus on the study of the dynamical clearing induced by a companion using both analytical considerations (Section \ref{analyticalcalc}) and hydro-dynamical simulations (Section \ref{hydro}). 
We note that the profiles used for the fitting procedure with DALI are based on simplified physical models which qualitatively represent the characteristic features of the real profile. As a consequence, performing hydrodynamical simulations becomes of paramount importance in order to verify that the DALI models can be qualitatively reproduced in hydrodynamical models.
This mechanism, indeed, leads to a hole/cavity empty of large grains which are filtered out, while the gas and small dust grains can still continue to be accreted by the central star \citep{Rice2006, Zhu2012}. 

\subsection{Analytical considerations}\label{analyticalcalc}
We explore the possibility that a massive planet located in the cavity region at a separation $R_{\rm p}\approx 15-25\; {\rm au}$  is responsible for the gas and dust density structure we observe. 

In order to open a gap in the gaseous disc, the planet mass $M_{\rm p}$ has to satisfy the following criterion 
\citep{crida2006}
\begin{equation}
\frac{3}{4}h_{\rm p}\left(\frac{M_{\rm p}}{3M_\star}\right)^{-1/3}+\frac{50 \nu_{\rm p}}{\Omega_{\rm p}R_p^2}\left(\frac{M_{\rm p}}{M_\star}\right)^{-1}\lesssim 1.\label{cridacrit}
\end{equation}
where $\nu$ is the viscosity parameter, $h_{\rm p}$ is the disc aspect-ratio at $R=R_{\rm p}$ (from here below, the subscript ``p'' indicates that the quantity is computed at the planet location) and $\Omega_{\rm p}\approx\sqrt{GM_\star/R_{\rm p}^3}$ is the standard Keplerian orbital frequency of the planet, where $G$ is the gravitational constant. 
Using the disc parameters obtained from the DALI models in the sections above and using a \citet{shakura73} prescription for the viscosity at the planet location $\nu_{\rm p}=\alpha_{\rm SS} h_{\rm p}^2\Omega_{\rm p}R_{\rm p}^2$, assuming a fiducial value of $\alpha_{\rm SS}=0.005$ (we motivate this choice in the next section), we obtain
\begin{equation}
M_{\rm p}\gtrsim 0.002\textrm{--}0.006 M_\star\approx 3\textrm{--}9\;M_{\rm J},\label{eq:mplan}
\end{equation}
where we have assumed $R_{\rm p}=20 \,{\rm au}$, as suggested by the location of the cavity edge in the gas, and explored a range of values of the disc thickness $h_{\rm p}$ consistent with $h_{\rm c}=0.07\textrm{--}0.12$ ($h_{\rm p}=h_{\rm p}(R_{\rm p}/R_0)^\psi$), which is poorly constrained by observations and constitute the source of uncertainty in Eq. (\ref{eq:mplan}). 
We notice that the minimum mass expected to open a gap is affected by our fiducial choice of $\alpha_{\rm SS}$. In particular, lower values of $\alpha_{\rm SS}$ would require lower planet masses ($M_{\rm p}\approx 0.3\, M_{\rm J}$ with $\alpha=10^{-4}$ and $h_{\rm p}=0.07$) in order to satisfy the criterion.

Assuming $M_{\rm p}=3\textrm{--}9\,M_{\rm J}$,
an estimate of the gap width $\Delta$ produced by such a planet can be obtained using \citet{lin1979}
\begin{equation}
 \Delta\approx\left[\left(\frac{M_{\rm p}}{M_\star}\right)^2\frac{\Omega_{\rm p} R_{\rm p}^2}{\nu_{\rm p}}\right]^{1/3}R_{\rm p}\approx 20 \;{\rm au}.
\end{equation}\label{eq:rcav}
The gas depletion factor provided by such a planet is given by \citep{kanagawa18}
\begin{equation}
\frac{\Sigma_{\rm min}}{\Sigma_0}=\frac{1}{1+0.04K}\approx 0.05\textrm{--}0.08
\end{equation}
where
\begin{equation}
K=q^2h_{\rm p}^{-5}\alpha_{\rm SS}^{-1}.
\end{equation}
We note that the condition to open a gap in the gas is a sufficient condition to open a gap also in the dust \citep{dipierro2017}.

The presence of the planet can reduce the mass flux across the planet orbit, depending on the disc thickness and planet-to-star mass ratio \citep{farris2014,young2015,ragusa2016}. If the planet migration rate is slower than the viscous spreading of the disc, such a ``dam'' effect induces a reduction of the disc density downstream of the planet orbit, producing as a consequence an extended cavity rather than a thin gap structure. 
Gap opening planets undergo the so-called Type II migration, which occurs on a timescale \citep{syer1995,Ivanov1999}
\begin{equation}
t_{\rm\, type II}\approx\frac{M_{\rm p}+M_{\rm d}^{\rm local}}{M_{\rm_d}^{\rm local}}\,t_\nu\approx 2\textrm{--}3.3 \,t_\nu \approx 4\textrm{--}6\times 10^3 t_{\rm orb},\label{eq:migtime}
\end{equation}
where $t_{\rm orb}=2\pi\Omega_{\rm p}^{-1}$ is the planet orbital period. In Eq. \ref{eq:migtime} we assumed $M_{\rm p}=3\textrm{--}9\,M_{\rm J}$ and $R_{\rm p}=20 \,{\rm au}$ and $M_{\rm d}^{\rm local}=4\pi\Sigma_{\rm p}R_{\rm p}^2\approx 4 \,M_{\rm J}$, that is approximately the amount of gas contained within the planet orbit, as estimated again based on the DALI modeling above. We thus expect that in the current situation a planet of $M_{\rm p}\gtrsim 3 M_{\rm J}$ should be able to produce a gap in the inner disc. Furthermore, the migration timescale being larger than the viscous time could in principle favour the depletion of the inner disc producing a cavity. 

\subsection{Numerical Simulations}\label{hydro}
We perform a set of 3D numerical simulations of a planet and a gas + dust accretion disc using the code \textsc{phantom} \citep{Price2017} in order to find the mass of the planet producing the density structure expected from the DALI modeling of the observations.

We use a locally isothermal equation of state: the gas temperature is a radial power-law that produces a disc thickness $H= h_{\rm c}(R/R_{\rm 0})^\psi R$ (where $\psi=0.05$, bold parameter in Table \ref{tab:dali_new}). 
The planet and the star are modeled by two sink particles \citep{Bate1995}. The planet is initially on a circular orbit and is allowed to accrete and migrate. The sink particles exert the standard gravitational force on gas and dust particles. The motion of the sinks is computed step by step from the force they exert on each other and from the back-reaction of the disc. Particles are considered accreted when their distance from the sink is below the sink radius (see next section), and they are gravitationally bound to it.

The profile of ``large'' dust grains in the DALI modeling describes the properties of a vast population of different grain sizes ($1 {\rm \mu m}\lesssim a_{\rm large}\lesssim 1 {\rm cm}$). In our simulations however, we do not follow a dust population with a grain size distribution, but we only model individual sizes. 

To describe the dynamics of dust grains we adopt the one fluid algorithm developed in \citet{Laibe2014}, \citet{Price2015} and \citet{ballabio2018} for 
large grains (for which we considered two different sizes: $a_{\rm large}=0.1 {\rm mm}$ and $a_{\rm large}=0.5 {\rm mm}$). All grain sizes we used are characterized by a Stokes Number ${\rm St}<1$ throughout the entire disc, so that the dust dynamics is correctly described by the one fluid SPH algorithm \citep{ballabio2018}.  

We performed some simulations using also an additional family of dust grains of size $a_{\rm small}=1\,\mu m$, with the aim to investigate the dynamics of small dust particles that are part of the modeling with DALI. We notice that, as expected and previously discussed in Sec. \ref{sec:dgradprof}, very small dust grains ($a_{\rm small}=1\,{\rm \mu m}$) reproduce closely the dynamics of the gas, being tightly coupled to it. 
As a consequence, we choose not to investigate the dynamics of $a_{\rm small}=1 \,{\rm \mu m}$ grains since the information about their distribution can be inferred by merely rescaling the gas profile with the appropriate dust-to-gas ratio. 

Angular momentum transfer, allowing gas accretion, is provided by SPH artificial viscosity. We choose the artificial viscosity parameter $\alpha_{\rm AV}=0.2$ in order to have an equivalent \citet{shakura73} turbulent parameter $\alpha_{\rm SS}\approx 0.005-0.01$ throughout the entire disc. Our choice of $\alpha_{\rm SS}$ is dictated by the minimum viscosity required in an SPH numerical simulation in order to properly provide shock resolution.

Each simulation is evolved for $t=60\, t_{\rm orb}$, which corresponds to $\sim$ 4500 years. 
We run all our simulations with $N_{\rm part}=7.5\times 10^5$ particles. 
We also run two simulations using a larger number of SPH particles ($N_{\rm part}=1.5\textrm{--}2.2\times 10^6$) in order to check the reliability of our results at lower resolution.
	
\subsubsection{Initial conditions}
\begin{table}
\begin{tabular}{lcccccc}
\hline
Ref.&$0.1\,{\rm mm}$&$0.5\,{\rm mm}$&$M_{\rm p}$&$R_{\rm p}$&$h_{\rm c}$&$N_{\rm part}$\\
\hline
1&x&&$3\,M_{\rm J}$&$20\,{\rm au}$&$0.07$&$7.5\times10^5$\\
2&x&&$6\,M_{\rm J}$&$20\,{\rm au}$&$0.10$&$7.5\times10^5$\\
3&x&&$9\,M_{\rm J}$&$20\,{\rm au}$&$0.12$&$7.5\times10^5$\\
4&&x&$3\,M_{\rm J}$&$20\,{\rm au}$&$0.07$&$7.5\times10^5$\\
5&&x&$6\,M_{\rm J}$&$20\,{\rm au}$&$0.10$&$7.5\times10^5$\\
6&&x&$9\,M_{\rm J}$&$20\,{\rm au}$&$0.12$&$7.5\times10^5$\\
\hline
7&x&&$9\,M_{\rm J}$&$20\,{\rm au}$&$0.12$&$1.5\times10^6$\\
8&x&&$9\,M_{\rm J}$&$20\,{\rm au}$&$0.12$&$2.2\times10^6$\\
\hline
\end{tabular}
\caption{Summary of the  numerical simulations. Ref. is the reference number of the simulation, $0.1\, {\rm mm}$ and $0.5\, {\rm mm}$ refer to the large dust grain size we used, $M_{\rm p}$ is the planet mass, $R_{\rm p}$ its separation from the central star, $h_{\rm c}$ is the disc aspect-ratio at $R_{\rm 0}=56 \, {\rm au}$ and finally $N_{\rm part}$ is the total number of particles. Simulations 7 and 8 are convergence tests that we performed with a larger number of particles.}\label{tab:sim}
\end{table}

The initial conditions of our simulations consist of a planet with mass ranging $M_{\rm p}=3\textrm{--} 9 M_{\rm J}$ at a separation from the central star of $R_{\rm p}=20\, {\rm au}$. 
The choice of the planet mass and its separation from the star has been made starting from the considerations in Eq. (\ref{eq:mplan}).

We run simulations using three different couples $(M_{\rm p},h_{\rm c})$ for which the gap opening criterion in Eq. (\ref{eq:mplan}) is satisfied.
We note that the disc aspect-ratio $h$ is poorly constrained from observations considering that it is strongly affected by the estimate of the luminosity of the star (as previously discussed).

The planet is initially completely embedded in the disc, that is assumed to have an unperturbed density profile (no cavity is present at the beginning of the simulation neither in the gas nor in the dust density profiles). The gas initial surface density profile uses the parameters obtained from the power-law DALI density model (bold parameters in Tab. \ref{tab:dali_new}, i.e. dashed red curve in Fig. \ref{fig:sigma}, respectively) without depletion. 
The large dust grains initial density profile uses for simplicity the self-similar DALI model introduced in Sec. \ref{selfsim_dust}, but using a tapering radius larger than $R_0$ ($R_{\rm taper}=70\,{\rm au}$): the steep profile at large radii observed in the large dust grain surface density profile from the DALI modeling is the natural outcome of large dust grains radial drift; as the simulation starts, the large grains drift inward and the dust density profile steepens. This demonstrates that the narrow dust ring observed with ALMA is most likely due to radial drift of large dust grains. 

It should be noted that thicker discs are more efficient in transporting the angular momentum throughout the disc. As a consequence, as can be seen in Eq. (\ref{cridacrit}), for a fixed mass of the planet, the gap opening in the gas becomes progressively less efficient as the disc aspect-ratio grows.
This, in fact, introduces a degeneracy for the couples of parameters $(M_{\rm p},h)$, that is confirmed by the results of our numerical simulations. 

In Table \ref{tab:sim} we report a summary of the simulations that better reproduce the profiles from the DALI modeling.
We note that our high resolution runs (simulations 7 and 8 in Table \ref{tab:sim}) show overall good agreement with lower resolution runs except in the very inner region of the disc (see the end of Sec. \ref{sec:results}).

\begin{figure*}[]
\begin{center}
\includegraphics[width=0.32\textwidth,angle=0]{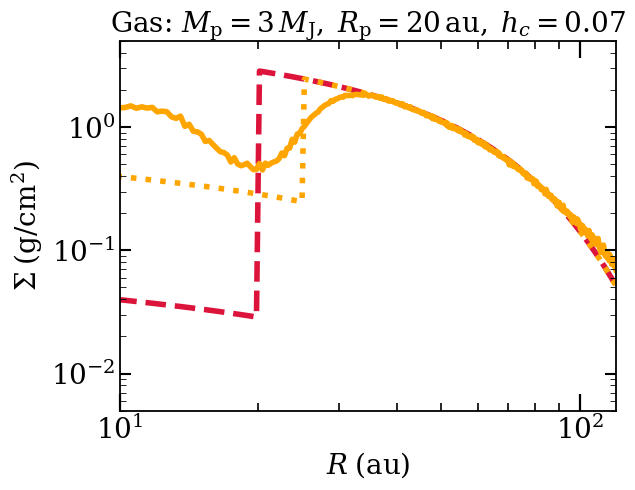}
\includegraphics[width=0.32\textwidth,angle=0]{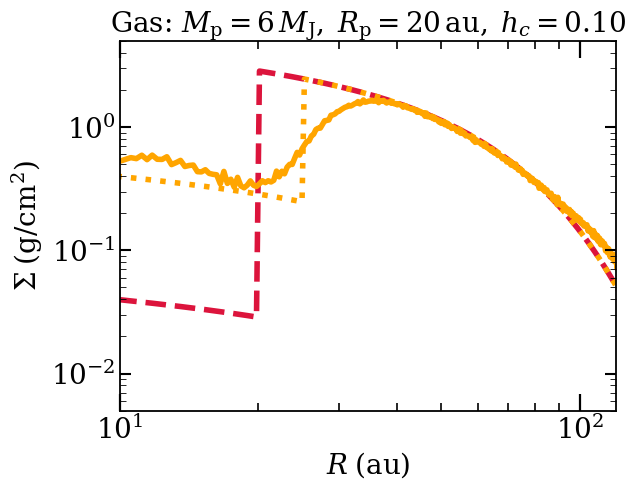}
\includegraphics[width=0.32\textwidth,angle=0]{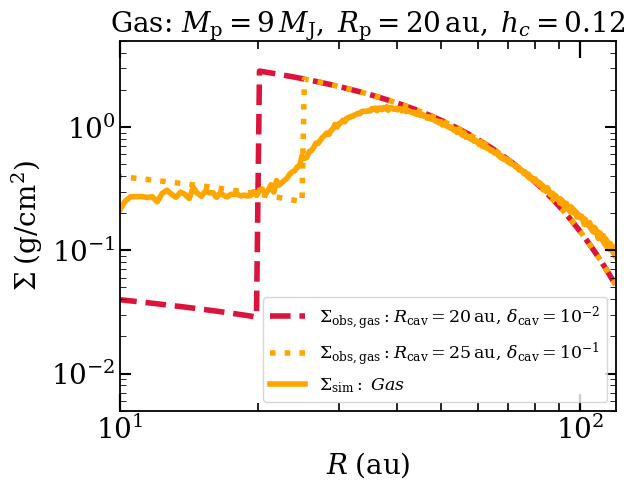}\\
\includegraphics[width=0.32\textwidth,angle=0]{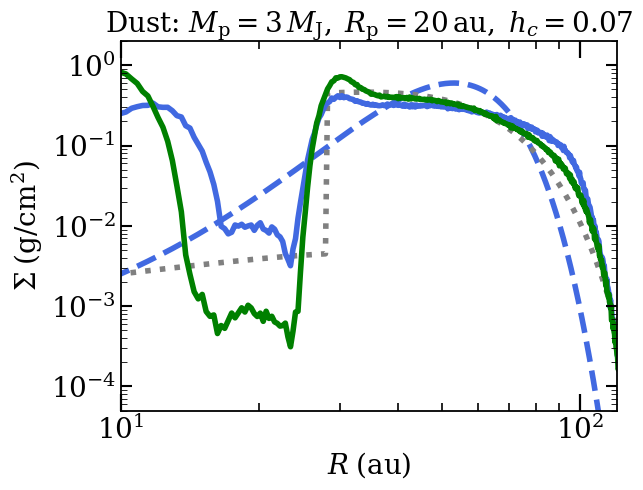}
\includegraphics[width=0.32\textwidth,angle=0]{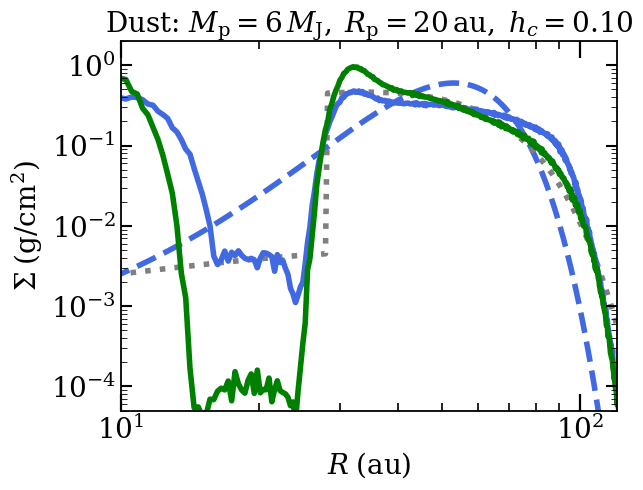}
\includegraphics[width=0.32\textwidth,angle=0]{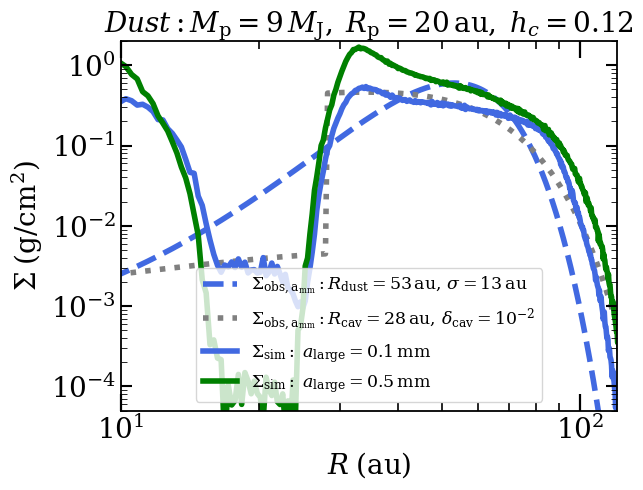}\\
\includegraphics[width=0.31\textwidth,angle=0]{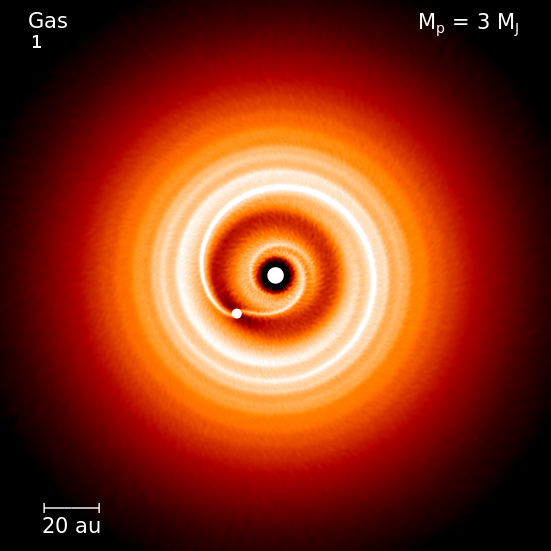}
\includegraphics[width=0.31\textwidth,angle=0]{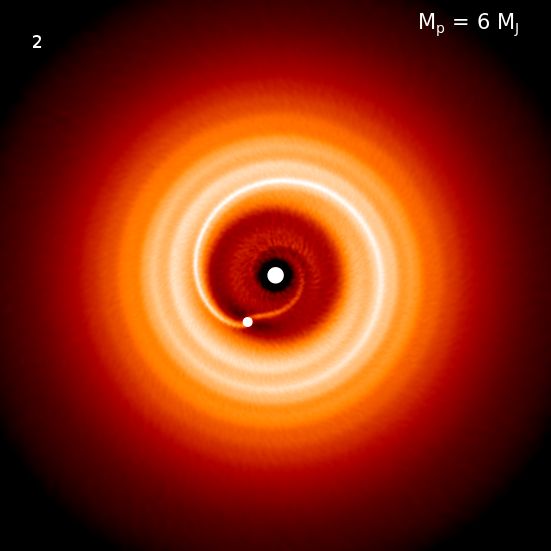}
\includegraphics[width=0.31\textwidth,angle=0]{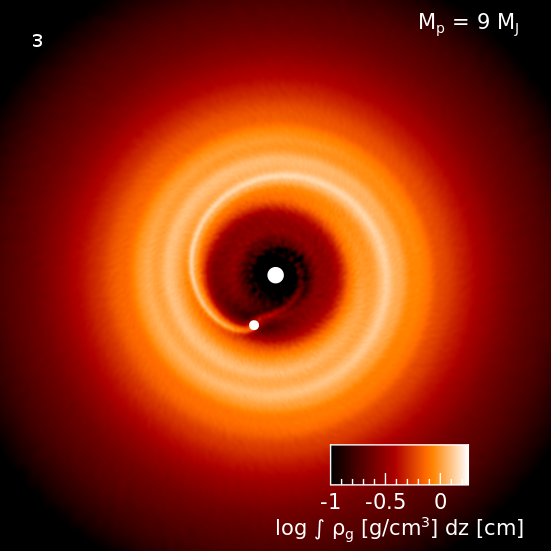}\\
\includegraphics[width=0.31\textwidth,angle=0]{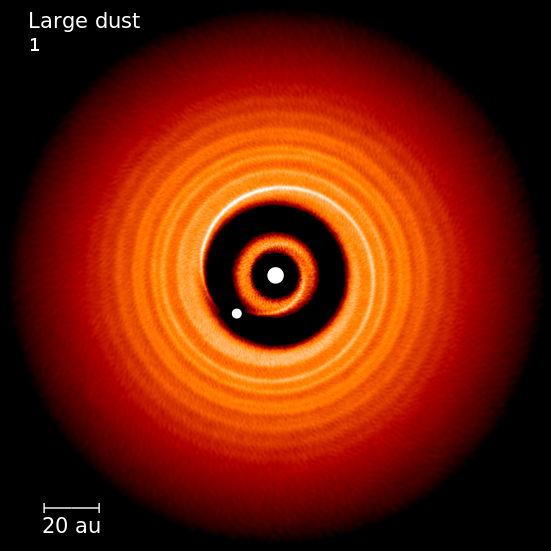}
\includegraphics[width=0.31\textwidth,angle=0]{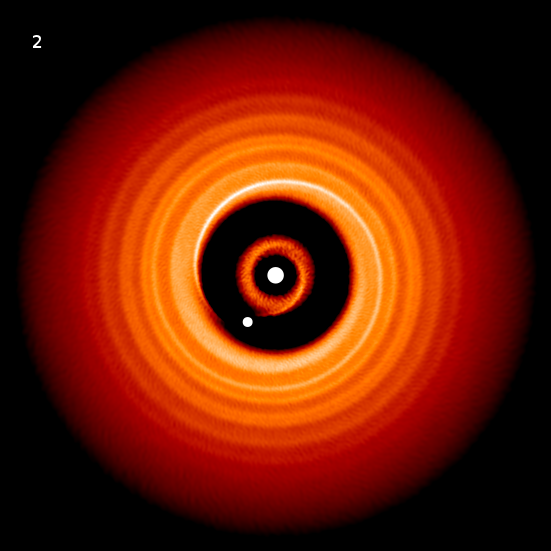}
\includegraphics[width=0.31\textwidth,angle=0]{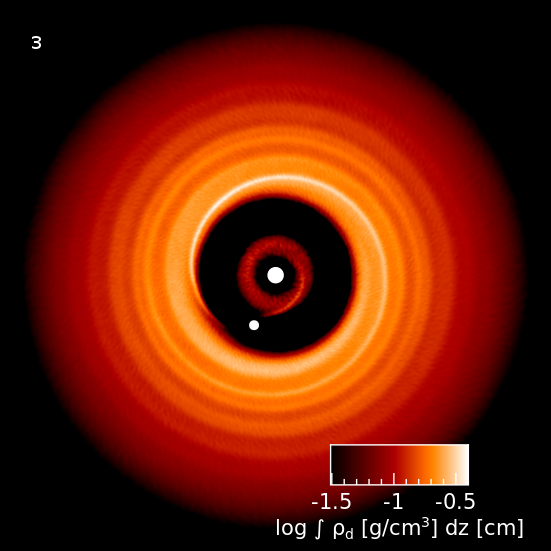}
\caption{Profiles (top row) and colour plots (central and bottom row) of the surface density of both gas and dust. Different columns show the results of simulations for different planet masses $M_{\rm p}$ and aspect ratios $h_{\rm c}$. 
The first row and second row show the azimuthally averaged surface density of gas and dust, respectively. In particular from our simulations of gas (orange solid curve), large dust grains  $a_{\rm large}=0.1 \,{\rm mm}$ (blue solid curve, from left to right simulation 1, 2, 3, respectively) and $a_{\rm large}=0.5 \,{\rm mm}$ (green solid curve, from left to right simulations 4, 5, 6, respectively) after $t\approx 60\,t_{\rm orb}$ of evolution, to be compared to the surface density profiles obtained from the DALI modeling for the gas (red dashed and orange dotted curves, producing the CO isotopologues emission in Fig. \ref{fig:fig7} with the same colours) and for the large dust grains (blue dashed and grey dotted curves producing the continuum emission profiles in Fig. \ref{fig:continuumvsR} with the same colours). Dashed curves represent the best denisty models from DALI, while dotted ones are those DALI models that best fit the simulations.
The third row shows the surface density colour-plot of the gas (from simulations 1, 2, 3, as indicated in the left top corner of the images).
The fourth row shows the surface density colour-plot of the large dust grains $a_{\rm large}=0.1 \,{\rm mm}$ (from simulations 1, 2, 3, as indicated in the left top corner of the images).
The colour-scale is logarithmic; the large white dot represents the central star and the small one shows the planet position.
} \label{fig:densimsnap}
\end{center}
\end{figure*}

\subsubsection{Results}\label{sec:results}
In all numerical simulations reported in Table \ref{tab:sim} the planet rapidly ($t\approx 15\, t_{\rm orb}$) carves a gap in the initially unperturbed disc (both in the gas and in the dust), as expected from the gap opening criterion presented in Eq. (\ref{eq:mplan}).

Fig. \ref{fig:densimsnap} shows the results of our numerical simulations, from left to right with $M_{\rm p}=$ 3, 6 and 9 $M_{\rm J}$, respectively, for the various species. 
The first and second row shows the azimuthally averaged density profiles for gas (orange solid lines), for 0.5 mm-sized dust (green solid lines), and for 0.1 mm-sized dust (blue solid lines). 
For a direct comparison, the same plots also show the surface density profiles obtained from the DALI modeling. The red dashed line is the reference model for the gas (boldface parameters in Table 2), which has a gas depletion factor of $\delta_{\rm gas}=10^{-2}$, while the orange dotted line is the same model but with a smaller gas depletion in the cavity, $\delta_{\rm gas}=10^{-1}$, and larger cavity size, $R_{\rm cav}=25\, {\rm au}$. These gas profiles produce the red and orange emission profiles in Fig. \ref{fig:fig7}, respectively. The blue dashed line shows the reference Gaussian ring density profile for the ``large dust grains'' obtained from DALI, which produces the blue curve in the continuum emission profile in Fig. \ref{fig:continuumvsR}; the grey dotted curve is the best-matching power-law density profile for the large dust grains obtained with DALI and discussed in Sec. 4.4, the dust continuum emission associated to it is the grey curve in Fig. \ref{fig:continuumvsR_pl}. The third and fourth rows of Fig. \ref{fig:densimsnap} show the surface density colour-plots of the corresponding simulations in the upper panels (Ref: 1, 2, and 3 in Table \ref{tab:sim}).

In Tab. \ref{tab:sim} we report also two convergence test runs using a larger number of particles and the same parameter choice as our fiducial model ($N_{\rm part}=1.5\times 10^6$ and  $N_{\rm part}=2.2\times 10^6$ simulations 7 and 8, respectively). These high resolution runs are in good agreement with the lower resolution one throughout the entire disc both for gas and dust, apart from the inner disc region where at low resolution the lower density produces an increase of SPH artificial viscosity, and thus a spurious fast evacuation of the innermost part of the cavity.

We finally note that the mass of the planet does not change significantly during our simulations. In particular, in all cases, we observe a growth of the planet mass at the end of the simulation of a fraction of $M_{\rm J}$. The correspondent growth rate at the end of the simulation is of $\dot M_{\rm p}\sim 5\times 10^{-5} \,{\rm M_{\rm J}\,yr^{-1}}$, consistent with the expected theoretical prediction for the planet accretion rate (see Eq. 15 of \citealp{dangelo2008}).

\subsubsection{Radiative transfer modelling}\label{sec:radmc3d}
\begin{figure}
\centering
\includegraphics[width=80 mm,angle=0]{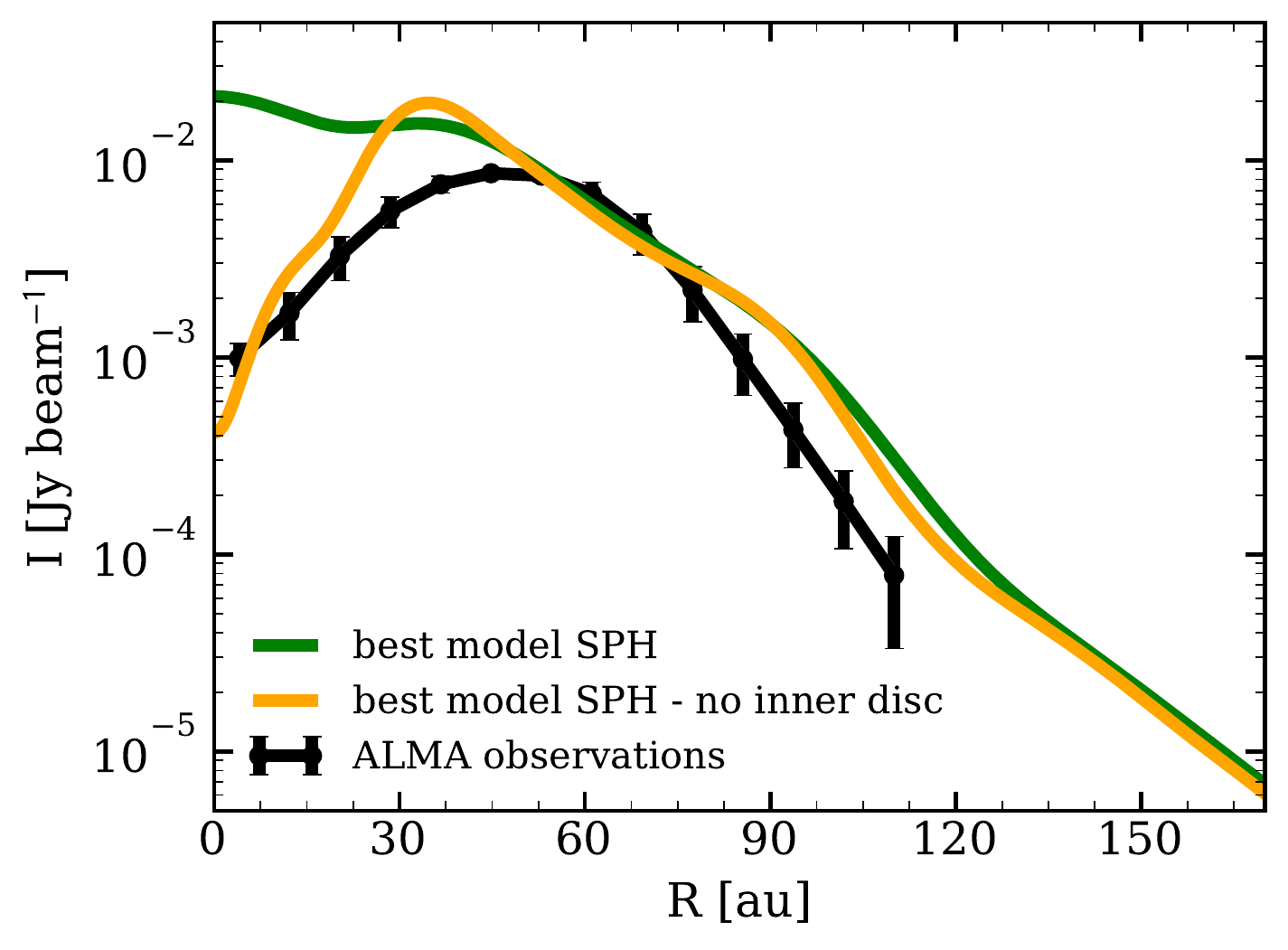}
\caption{Radial profiles of the dust component of the hydrodinamical simulations after radiative transfer modelling. The black dots are our data in the continuum (1.3~mm); the green and yellow lines are respectively the model with and without the presence of the inner disc. The best model here listed is using R$_{\rm p}$ = 20~au, M$_{\rm comp}$ = 6 M$_{\rm Jup}$. The size of large grains is taken as 0.1~mm.} \label{fig:radmc3d}
\end{figure}

We have processed our hydrodinamical model results using a radiative transfer tool (RADMC3D, \citealt[][]{RADMC3D}) for the dust component, in order to test if our SPH model is qualitatively consistent with the observations. The images obtained were, then, convolved with a gaussian beam of $0.\!\!^{\prime\prime}15$ as done in DALI and compared with the observational data. In particular, we found the best agreement with the data in the model with the planet at R$_{\rm p}$ = 20~au, M$_{\rm comp}$ = 6 M$_{\rm Jup}$ and the size of large grains of 0.1~mm (see Fig. \ref{fig:radmc3d}).
We show the observations at 1.3 mm (black line) along with the synthetic dust flux profiles convolved based on radiative transfer calculation. The plots show the model presented in the Section \ref{sec:results} in green considering also the inner disc component; the model in yellow, instead, does not take into account the presence of inner disc and uses a cavity lowered of the same factor from the planetary radius up to the sublimation radius. Our simulation (both the yellow and green lines) well describe the outer disc, which reaches a semi-stable state in few dynamical timescales. Whereas, the inner disc is not well reproduced by the hydrodynamical simulations. Its flux is, indeed, higher than the one of observations, leading to a disc where the carved cavity is not well seen (see Fig. \ref{fig:radmc3d}). We expect, however, the inner disc within the planetary orbit to drain with time on a longer timescale than the outer disc and this might affect the dust profile in the inner regions. With a lower inner disc emission, we would be able to reproduce qualitatively the depth and separation of the planet (yellow line in Fig. \ref{fig:radmc3d}), main goal of hydro modelling we provided. 
A more detailed analysis on the inner disc will be addressed in future works.

\section{Discussion}\label{6}
We here investigate if a planet could be responsible for the formation of a cavity in the gas and dust profile similar to the one found through the chemical-physical code DALI. We note the following:
\begin{enumerate}
\item A planet mass of the order of $3M_{\rm J}$ does not reproduce the strong gas depletion in the inner disc inferred from the thermo-chemical modeling with DALI. On the other hand, a larger planet mass ($M_{\rm ]}\approx 6-9M_{\rm J}$) does a much better job, although the $6M_{\rm J}$ case requires a slightly thinner disc ($h=0.1$) than that obtained from the DALI modeling.

\item Overall, we find both our gas and dust density profiles qualitatively reproduce the ones inferred from thermo-chemical modeling of the observed fluxes. However, we note two issues. Firstly, concerning the gas, the best representative DALI models predict a gas depletion factor of the order of $10^{-2}$, which is not achieved in our simulations, that only show a gas depletion of $10^{-1}$ (see red and orange curves in CO isotopologues DALI model Fig. \ref{fig:fig7} and simulations Fig. \ref{fig:densimsnap}). Such a high gas depletion is essentially required by the observed gap in $^{13}$CO. A higher gas depletion might be achieved using smaller values of $\alpha_{\rm SS}$ and $h_c$, or much larger values of the planet mass. The latter possibility is unlikely at this planetary location, since it would imply that the planet mass falls within the brown dwarf mass regime. Direct imaging observations, indeed, ruled out the presence of a companion with mass higher than 10 M$_{\rm Jup}$ beyond 20 au (Benisty et al. 2019, in prep.).

\item Concerning the dust, our hydrodynamical model involving one planet reproduces well the ring feature at $\approx 50$ au, but not the inner region of the cavity, where an inner dust disc close to the star still remains. While a semi-stationary state have been reached outside the planet orbit (i.e. we do not expect the gap edge location, depletion at the planet location and outer density profile to change significantly at longer timescales), we note that the dust in the inner disc decouples from the gas and is not effectively accreted onto the central star along with the gas. This implies that further depletion of the dust in the inner cavity might occur at later times. 

\item
Good agreement between the outcome of our simulations and the DALI gas density profile can be obtained with the model having $\delta_{\rm gas}=10^{-1},\, R_{\rm cav}=25$; in particular, we note that the location of the cavity edge for the gas in this model (``orange'' CO isotopologues DALI model in Fig. \ref{fig:fig7}) is perfectly consistent with the density structure provided by a planet placed at $R_{\rm p}=20\,{\rm au}$ in our simulations.
This DALI model, despite not being the best representative for the gas, reproduces fairly well the fluxes for $^{12}$CO and C$^{18}$O, but it fails in reproducing the $^{13}$CO.

\item
The dust profiles from the numerical simulations appear to best reproduce the DALI self-similar model (Section \ref{selfsim_dust}) with $\delta_{\rm dust}=10^{-2}$ and $R_{\rm cav}=28$ au (grey model in Fig. \ref{fig:continuumvsR_pl} and \ref{fig:densimsnap}). The hydrodynamic density profiles, in particular those with  $a_{\rm large}=0.1$ mm, reproduce the location of the cavity edge and depth of the large dust grains density distribution of this model (see blue solid curves compared to grey dotted ones in Fig. \ref{fig:densimsnap}). 
This does not occur for larger grains ($a_{\rm large}=0.5\, {\rm mm}$) that instead produce a deeper gap with respect to that in the modeling, as shown in the second row of Fig. \ref{fig:densimsnap}. 

\item Our hydrodynamical simulations are not able to reproduce the Gaussian density profile of the large dust grains that provides the best match with the continuum flux (blue curves in Fig. \ref{fig:continuumvsR} and \ref{fig:densimsnap}). We note that, in general, any planet model would predict a sharp disc truncation beyond the planet location instead of a shallow Gaussian depletion. As we show in Section \ref{sec:radmc3d}, a sharp density feature is smoothed when synthetic dust flux  profile have been created and then convolved. The results found can qualitatively describe the observations as well as the gaussian profile if the inner disc present in the simulations is neglected. 

\item
We find that the best match between our hydrodynamical simulations and DALI density profiles is provided by grains with $a_{\rm large}=0.1$ mm with respect to those with $a_{\rm large}=0.5$ mm; this suggests that the dynamics of the entire large dust grain population can be best described by the smaller ($\approx 0.1$ mm) sizes, rather than the larger ones ($\approx 0.5$ mm).  

A more accurate treatment would require a multi-grain approach \citep{hutchinson2018,dipierro18}, in order to account for the different dynamics of different grain sizes characterized by different Stokes numbers. 
\end{enumerate}

\subsection{Comparison with other possible scenarios}\label{7}
We here briefly discuss our results to explain the cavity together with other possible mechanisms:\\
\\
\textbf{Clearing mechanism. }We find that a planet $M_{\rm p}=9\,M_{\rm J}$ located at $R_{\rm p}=20\,{\rm au}$ from the central star qualitatively (but not perfectly) reproduces the DALI modeling of gas and the large dust density distribution.  
The best match is obtained with $0.1\, {\rm mm}$ grains, implying that large dust grains behave qualitatively as fluid composed by grains with that size. In particular, good agreement is obtained at large radii, where the simulations clearly show that the planet might produce a ring-like feature in the dust distribution. However, we were not able to reproduce the dust depletion, expected from the DALI modeling, in the inner cavity using a one planet model, and at the end of our simulations we are still left with an inner disc of dust around the star. 
\\
\\
\textbf{Photoevaporation. }
Models of photoevaporation predict small cavities ($\lesssim$ 10~au) and low accretion rates. The debate about what is the driving mechanism of photoevaporation between FUV, EUV, X-ray and what are the exact values of mass-loss rate is still open. In order for the photoevaporation to start the accretion rate have to drop below the photoevaporation rate. As soon as it starts, a cavity free from gas and dust is generated, with the very large grains rapidly migrating inward \citep{Alexander2007}.
CQ Tau was considered by \citet{Donehew2011} and \citet{Mendigutia2012} to have an high accretion rate ($\dot{M}_{\rm acc}\lesssim 10^{-7}~M_{\odot}$ yr$^{-1}$), which can be considered high for disc dispersal to be the main mechanism acting in the formation of the CQ Tau cavity \citep{Owen2016}. A combination between the clearing mechanism and photoevaporation can, instead, be considered as a possible option (e.g. \citealt[][]{Williams2011, Rosotti2013, Rosotti2015}) to be futher investigated.\\
\\
\textbf{Dead zones. }The suppression of magnetorotational instability inside the disc can create regions of low disc ionization, the so called ``dead zones" (e.g. \citealt[][]{Flock2012}). In these regions the rate of gas flow decreases with an accumulation of gas in the outer edge. At the same time the pressure maxima is able to trap large particles with a consequent ring-like feature in both gas and dust components \citep{2015A&A...574A..68F}. The ring is in general located at about the same radius \citep{Pinilla2016} in both the components. However, when dead zones and MHD wind are present at the same time a difference in the inner radii position can be observed. The presence of dead zones can increase the turbulence present in the disc and generate asymmetric vortices in the disc \citep{Ruge2016} at (sub)mm wavelengths. Such structures can be also created by disc-planet interaction if the $\alpha$ viscosity is low. We cannot exclude this mechanism by our observations. Higher resolution data are need to better constrain the profile of the ring-like disc, mainly its symmetry and variability \citep{Pinilla2016}. However, dead zones can explain ring-like structures, but not cavities such as the one clearly present in the CQ Tau system. 
\\
\\
A combination between the mechanism described above can be also a possible solution which, however, we will not address in this paper.

\section{Conclusions}\label{8}
Discs where it is possible to clearly distinguish a cavity are of particular interest in the study of protoplanetary disc evolution and of the planet formation process. In particular, we have studied the disc around the CQ Tau pre-main-sequence star.
We have made use of spatially resolved ALMA observations of the dust continuum and of three CO isotopologues, $^{12}$CO, $^{13}$CO and C$^{18}$O, together with the observed SED of CQ Tau. We employed the chemical-physical code DALI to model the CQ Tau disc and to derive the surface density profile of the dust (small and large grains separately) and of the gas. Finally, we ran 3D SPH simulations to test whether an embedded planet could be able to create the observed disc structure. The main results from our analysis are summarized here:
\begin{itemize} 
\item In the CQ Tau disc the gas radial extent is a factor of two broader than that one of the large dust grains;
\item  CQ Tau shows a clear cavity in both gas and dust, that can be described by different functional forms. Assuming that the mm-sized grains are distributed in a Gaussian ring, the Gaussian peak position is located at a distance of 53 au from the central star and the Gaussian width is $\sigma=13\,$au. The cavity is present also in the gas component. The cavity radius and level of depletion are degenerate, but they can be constrained with reasonable uncertainties: the radius is smaller than 25 au and larger than 15 au and the depletion has to be between $10^{-1}$ and $10^{-3}$ with respect to the profile of a full self-similar disc. 
\item The computed dust-to-gas ratio is not radially constant throughout the disc. Moreover, the global dust-to-gas ratio is found to be $\sim 0.09$, higher than the typical values assumed ($\sim 0.01$) and most likely because of carbon depletion.
\item We performed a set of SPH simulations in order to investigate whether the presence of a massive planet might be responsible for the radial structure of the CQ Tau disc. We find that a massive planet with a mass of at least $M_{\rm p}=6 - 9\, M_{\rm J}$ located at $R_{\rm p}=20\,{\rm au}$ from the central star can reproduce well the ring feature at $\approx 50$ au but not the dust distribution in the inner cavity region of the best representative density model predicted using DALI. Also, our simulations do not reproduce the strong gas depletion within the cavity that appears from the best fit DALI modeling, although we caution the reader of possible degeneracies in both the thermo-chemical modeling (see above) and in the hydrodynamical modeling (for which a lower viscosity in the disc might reconcile the results). 

\end{itemize}

\section*{Acknowledgements}
We thank the anonymous referee for providing insightful comments. We are thankful to Antonella Natta, Antonio Garufi, Mario van den Ancker and Carlo F. Manara for useful discussions. This paper makes use of the following ALMA data: ADS/JAO.ALMA\#2013.1.00498.S, ADS/JAO.ALMA\#2016.A.000026.S, and ADS/JAO.ALMA\#2017.1.01404.S. ALMA is a partnership of ESO (representing its member states), NSF (USA) and NINS (Japan), together with NRC (Canada) and NSC and ASIAA (Taiwan) and KASI (Republic of Korea), in cooperation with the Republic of Chile. The Joint ALMA Observatory is operated by ESO, auI/NRAO and NAOJ. This work was partly supported by the Italian Ministero dell\'\,Istruzione, Universit\`a e Ricerca through the grant Progetti Premiali 2012 -- iALMA (CUP C52I13000140001), by the Deutsche Forschungs-gemeinschaft (DFG, German Research Foundation) - Ref no. FOR 2634/1 TE 1024/1-1, and by the DFG cluster of excellence Origin and Structure of the Universe (\href{http://www.universe-cluster.de}{www.universe-cluster.de}). AM and SF acknowledge an ESO Fellowship. GD and ER acknowledge financial support from the European Research Council (ERC) under the European Union's Horizon 2020 research and innovation programme (grant agreement No 681601). MT has been supported by the DISCSIM project, grant agreement 341137 funded by the European Research Council under ERC-2013-ADG and by the UK Science and Technology research Council (STFC). GL, MGUG, AM, SF, LT, LP and MT have received funding from the European Union’s Horizon 2020 research and innovation programme under the Marie Skłodowska-Curie grant agreement No 823823 (RISE DUSTBUSTERS project).



\bibliographystyle{mnras}
\bibliography{references.bib} 




\appendix

\section{SED}
Table \ref{tab:sed1} lists the fluxes of CQ Tau available in the literature between $360\,$nm and $3.57\,$cm.

\begin{table}
	\centering
	\caption{Values of the flux observation at different wavelengths. 1. \citet{Creech2002}, \citet{Mannings1997}; 2. \citet{Banzatti2011},  \citet{Testi2003}; 3.  \citet{Thi2001}; 4. \citet{Oja1987}; 5. \citet{Cutri2003}; 6. \citet{Mendigutia2012}}
	\label{tab:sed1}
	\begin{tabular}{||llr||} 
		\hline
		Wavelength & Flux & Reference\\
		$[\micron]$ & [Jy] & \\
		\hline
		0.36 & 0.055 & 4\\
		0.43 & 0.201 & 4\\
		0.55 & 0.378 & 4\\
		0.7 & 0.169 & 5\\
		0.9 & 1.23 & 6\\
		1.235 & 1.08 & 5\\
		1.66 & 1.54 & 5\\
		2.16 & 2.28 & 5\\
		3.4 & 2.6 & 3\\
		6.9 & 2.1 & 3\\
		9.6 & 7.1 & 3\\
		12 & 4.4 & 1\\
		17 & 13.1 & 3\\
		25 & 14.7 & 1\\
		28.2 & 21.6 & 3\\
		60 & 16.58 & 1\\
	 	100 & 12.52 & 1\\
	 	870 & 0.421 & 2\\
	 	1.3E+3 & 0.103 & 2\\
	 	2.7E+3 & 22E-3 & 2\\
	 	3.4E+3 & 13.1E-3 & 2\\
	 	6.9E+3 & 2E-3 & 2\\
	 	13.4E+3 & 2.8E-4 & 2\\
	 	35.7E+3 & 7.9E-5 & 2\\
		\hline
        \hline
	\end{tabular}
\end{table}


\bsp	
\label{lastpage}
\end{document}